\documentclass[sigconf,10pt]{acmart}
\renewcommand\footnotetextcopyrightpermission[1]{} 
\usepackage{algorithm}
\usepackage{algorithmic}
\usepackage{subfigure}
\usepackage{textcomp}
\usepackage{stfloats}
\usepackage{graphicx}
\usepackage{multirow}
\usepackage{array}

\newcommand{\PreserveBackslash}[1]{\let\temp=\\#1\let\\=\temp}
\newcolumntype{C}[1]{>{\PreserveBackslash\centering}p{#1}}
\newcolumntype{R}[1]{>{\PreserveBackslash\raggedleft}p{#1}}
\newcolumntype{L}[1]{>{\PreserveBackslash\raggedright}p{#1}}
\AtBeginDocument{
 \providecommand\BibTeX{{%
 \normalfont B\kern-0.5em{\scshape i\kern-0.25em b}\kern-0.8em\TeX}}}
\usepackage{enumitem}
\setitemize[1]{itemsep=0pt,partopsep=0pt,parsep=\parskip,topsep=0pt}
\setcopyright{none}

\acmConference[Mobicom 2020]{Mobicom 2020:The 26th Annual International Conference on Mobile Computing and Networking }{21-25 Sep,2020}{London, United Kingdom}

\begin{document}
\title{A Framework for Behavior Privacy Preserving in Radio Frequency Signal}

\author{Jianwei Liu}
\affiliation{%
  \institution{Zhejiang University}
  \city{Hangzhou}
  \country{China}}
\email{liujianwei@stu.xjtu.edu.cn}

\author{Jinsong Han}
\affiliation{%
  \institution{Zhejiang University}
  \city{Hangzhou}
  \country{China}}
\email{hanjinsong@zju.edu.cn}

\author{Lei Yang}
\affiliation{%
  \institution{The Hong Kong Polytechnic University}
  \city{Hong Kong}
  \country{China}}
\email{young@tagsys.org}

\author{Fei Wang}
\affiliation{%
 \institution{Zhejiang University}
 \city{Hangzhou}
 \country{China}}
\email{feiw2.ri@gmail.com}

\author{Feng Lin}
\affiliation{%
  \institution{Zhejiang University}
  \city{Hangzhou}
  \country{China}}
\email{flin@zju.edu.cn}

\author{Kui Ren}
\affiliation{%
  \institution{Zhejiang University}
  \city{Hangzhou}
\country{China}}
\email{kuiren@zju.edu.cn}

\setlength{\abovecaptionskip}{10pt} 
\setlength{\belowcaptionskip}{-10pt}

\begin{abstract}
Recent years have witnessed the bloom development of the human-centered wireless sensing applications, in which some human information, such as the user’s identity and motions, can be retrieved through analyzing the signal distortion caused by the target person. However, the openness of wireless transmission raises increasing concerns on user privacy, since either the human identity or human motion is sensitive in certain scenarios, including personal residence, laboratory and office. Researchers have reported that commodity WiFi signals can be abused to identify users. To dispel this threat, in this paper we propose a privacy-preserving framework to effectively hide the information of user behaviors in wireless signals while retaining the ability of user authentication. The core of our framework is a novel \textit{Siamese network}-based deep model, namely RFBP-Net.  In this way, wireless sensing reveals the user information moderately.  We conduct extensive experiments on both the real WiFi and RFID systems, and open datasets. The experiment results show that RFBP-Net is able to significantly reduce the activity recognition accuracy, i.e., 70\% reduction in the RFID system and 80\% reduction in the WiFi system, with a slight penalty in the user authentication accuracy, i.e., only 5\% and 1\% decrease in the RFID and WiFi system, respectively. 
\end{abstract}





\maketitle

\section{introduction}
Recently, wireless sensing techniques are promising and attractive  to retrieve information, such as the user’s identity and activity/gesture, in a non-intrusive and human-centered way \cite{DBLP:conf/infocom/ZhaoLLDHXG19,DBLP:journals/imwut/WangZ18,DBLP:journals/imwut/FanGL18,DBLP:journals/corr/abs-1810-04106,DBLP:conf/globecom/XinGWLYZ16,DBLP:conf/ipsn/ZengPM16,DBLP:conf/dcoss/ZhangWHK16,DBLP:conf/huc/0002LS16,DBLP:conf/mobihoc/ShiLLC17,DBLP:conf/globecom/LvYMDYG17,DBLP:conf/iecon/PokkunuruJBWS18}. The insight behind wireless sensing is that the RF signal can be distorted by human bodies during its transmission, with the effect of reflection, deflection, penetration, and the like. The wireless sensing enabled applications bring users convenient services, since even the pose 
\cite{DBLP:conf/cvpr/ZhaoLATZ0K18} and motion \cite{DBLP:journals/corr/abs-1904-00276} can be detected and identified in a fine-grained way. Therefore, the users' behaviors can also be extracted from the RF signals \cite{DBLP:journals/imwut/WangZ18,DBLP:conf/sensys/DingSYHZYXZ15,DBLP:journals/imwut/FanGL18,DBLP:conf/sensys/LiZMSB16,wang2014e-eyes:,wang2015understanding}.

\par However, the ability of sensing behaviors result in serious privacy concerns. Recently, Zhu \textit{et al.} \cite{DBLP:journals/corr/abs-1810-10109} have demonstrated that users' in-door coarse-grained behavior privacy, e.g., the presence of users, can be sensed by attackers through the WiFi signal analysis. In such cases, if users enjoy the RF-based service provided by a non-trusted service provider (SP), information leakage of their behaviors, sometimes sensitive, occurs. The SP may utilize collected signals to monitor the users and analyze their behaviors while the users are unaware of such surveillance. We argue  that this is a severe privacy threat because the SP like attacker can secretly supervise users and speculate about users' professions, interests and even the users’ password. For example, since 2013 the Carbanak gang has stolen tens of millions of dollars from several banks by hacking their IoT cameras and monitoring the clerks’ operations on the banking system \cite{robber}. 

\par We define such privacy as behavior privacy in RF signals
(RFBP). To achieve RFBP preserving, we aim to design a framework which can be controlled by clients to protect their poses, gestures, and activities in RF-based sensing applications. Meanwhile, the framework should retain RF-based applications’ normal functions as well as their performance, such as authentication, tracking and localization. In this paper, we chose the authentication as the typical and representative function deserving the retaining.

\par The core functionality of the framework is to filter out behavior features (\textit{i.e.}, behavior privacy) from the authentication-oriented data (AOD) while not destroying identity-relevant features. In this way, the behavior privacy-protected AOD can only be used to identify users. To achieve this aim, we should address the following challenges. First, RF signals contain both behavior information and identity information and it is difficult to figure out which part of the RF signal represents the behavior of the user and to filter them out. Second, we must not destroy the identity-relevant feature when excluding users' behavior information. In this paper, we overcome these challenges by skillfully converting this privacy preserving issue into a pure feature extraction issue. In the new issue, we aim to extract only identity-relevant pure feature while ignoring behavior-relevant feature from AOD. We achieve this goal by devising a novel Siamese network \cite{DBLP:conf/cvpr/ChopraHL05} based framework, whose core is RFBP-Net.

\par Basically, our framework first confirms that behavior privacy is contained in the AOD by using several classic learning models. Then a training set, in which each training sample has two kinds of labels (a behavior label and an identity label), is formed by using a well-designed algorithm. Afterwards, RFBP-Net is trained and all privacy-sufficient samples are processed by RFBP-Net. Finally, our framework utilizes several classic learning models to confirm the validity of the processed data.

\par In the experiment part, we first used a RFID system to evaluate our framework. Five volunteers were required to write ten different numbers from `0' to `9' in front of a tag array. Thus each signal sample had two labels: an identity label and an activity label. Then the collected signal samples were reconstructed and relabeled so that each training sample had two new labels: a similarity label and an identity label. We tested privacy-preserved data with different learning models and the experiment results demonstrated that by using extracted features, i.e., the processed dataset, the accuracy of activity recognition decreased from 95\% to 25\% while the accuracy of identity authentication only dropped 5\%. In the second experiment, we collected gesture based RF signals with ten volunteers in a WiFi system. Each volunteer posed ten different gestures to represent ten different behaviors. The evaluation results showed that RFBP-Net efficient protected gesture privacy while causing almost zero reduction in the identity authentication accuracy. In extensive experiments, we evaluated our framework with two open datasets \textit{Wiar} and \textit{Widar3.0}. The results demonstrated that our framework was significantly efficient at behavior privacy preserving.  

\par In summary, our contributions are threefold: 
\begin{itemize}
    \item We first notice that users' behavior privacy may be filched in RF signal-based applications, and then propose the concept of RFBP.
    \item We propose a novel framework, which aims to protect behavior privacy by using a novel deep model named RFBP-Net.
    \item We built a prototype of RFBP-Net and evaluated our framework with a RFID system and a WiFi system. The experiment results show that RFBP-Net performs well at behavior privacy preserving. The extensive experiments on two open datasets also prove that our framework delivers outstanding performance.
\end{itemize}

\par Of the remaining sections of this paper, Section \ref{section:related_work} is used to introduce existing related works. Section \ref{setion:background} first introduces the background to RF signal-based user authentication and activity/gesture recognition techniques and then introduces the function of the \textit{Siamese network}. The methodology is presented in Section \ref{sec:methodology}. All of the experiments and corresponding results are arranged in Section \ref{sec:experiment}, Section \ref{experiment_wifi} and Section \ref{sec:open_dataset}. In Section \ref{sec:discussion}, we discuss some issues and future works. In Section \ref{conclusion}, we conclude this paper.  

\section{Related Work}
\label{section:related_work}
In this section, we first introduce some RF signal-based authentication methods and then introduce several RF signal-based activity/gesture recognition systems.

\par \textbf{RF signal-based user authentication: }a wealth of recent works focused on authentication using RF signals. For user authentication, RF-Mehndi \cite{DBLP:conf/infocom/ZhaoLLDHXG19} leverages the coupling effect to amplify the variety of signal phase caused by the hand's impedance while touching. WiPIN \cite{DBLP:journals/corr/abs-1810-04106} extracts the body features from WiFi signals after propagating through the human body to authenticate users. FreeSense \cite{DBLP:conf/globecom/XinGWLYZ16} conducts user identification in in-door environments with WiFi signals. WiWho \cite{DBLP:conf/ipsn/ZengPM16}, WiFiU\cite{DBLP:conf/huc/0002LS16} and WiFi-ID \cite{DBLP:conf/dcoss/ZhangWHK16} made a authentication scheme that uses users' walking patterns to identify them. 

\par Our framework is different from existing RF signal-based authentication works. Previous works only focus on the authentication accuracy and user-friendliness instead of privacy protection. However, our framework can identify the identity of the user accurately, but the behavior privacy protection is also guaranteed.

\begin{figure*}[t]
    \centering
    \includegraphics[scale=0.7,trim=0 435 5 5,clip]{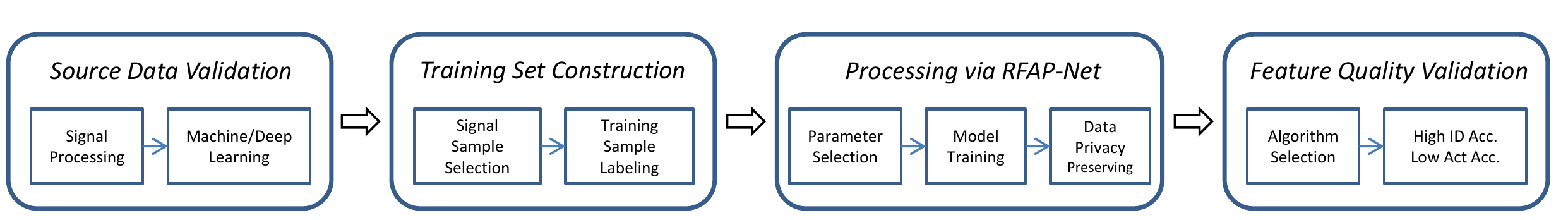}
    
    \caption{RFBP-Net consists of four modules: \textit{source data validation}, \textit{Training set construction}, \textit{processig via RFBP-Net} and \textit{feature quality validation}}
    \label{fig:workflow}
\end{figure*}

\par \textbf{RF signal-based activity/gesture recognition: }Activity/gesture recognition techniques are usually achieved by extracting behavior-relevant features from signal indicators in RFID signals or WiFi signals. By using RFID system, TACT \cite{DBLP:journals/imwut/WangZ18} first model the RF signals' intrinsic characteristics generated in application scenarios and then recognizes activities based on the model through phase analysis. Likewise, TagFree \cite{DBLP:journals/imwut/FanGL18}, which also uses RFID system as a foundation, recognizes activities by analyzing the signals distorted by multi-path. Li \textit{et al.} \cite{DBLP:conf/sensys/LiZMSB16} realize an activity recognition system under the light of deep learning. Through fine-grained signature extraction, E-eyes \cite{wang2014e-eyes:} identifies activities in WiFi settings. Wang \textit{et al.} \cite{wang2017device-free} also proposed a activity recognition system using device-free WiFi devices. A survey on WiFi-based activity recognition systems was organized in \cite{guolinlin_1}.
 
 \par As distinct from above mentioned related works, our framework aims to hide users' behavior information rather than to extract and utilize it. 

\section{Preliminary}
\label{setion:background}
A brief introduction to RF signal-based user authentication and activity/gesture recognition is given in the first and second part of this section. The third part is used to introduce the basic function of the \textit{Siamese network}. 

\subsection{RF signal-based user authentication}
RF signals such as RFID signals and WiFi signals are ubiquitously employed to authenticate users. The signal indicators utilized for feature extraction are the signals' RSS and phase. RSS depicts the strength of the signal, the value of RSS varies according to multiple environment variables, e.g., the traveling distance and the electric permittivity of the traveling media. The impedance of the human body, which is a kind of electric characteristic, varies among different individuals. Thus different individuals would cause different strength losses while signals are propagating through their bodies. Hence, RSS can be used for user authentication. The phase of the RF signal is denoted as:
\begin{equation}
\theta = (2\pi \dfrac{2d}{\lambda}+ {\theta}_i) \mod{2\pi},
\end{equation}
where $d$ is the propagation distance. The initial phase and the wavelength are denoted as $\theta_i$ and $\lambda$, respectively. During traveling and penetrating, $d$ is influenced by the motion and thickness of the body tissue accordingly. Therefore, phase is one of the common indicators that can represent the identity features of users.  

\subsection{RF signal-based activity/gesture recognition}
The principle of the activity/gesture recognition is similar to the principle of user authentication, i.e., the common signal indicators used for activity recognition generally are RSS and phase as well. The feasibility behind the activity/gesture recognition is the multi-path effect. The signal's traveling path is determined not only by linear propagation distance but also by the reflection and refraction caused by human bodies. Once the propagating signal is obstructed by the human body, the propagation path changes, resulting in the changes of signal indicators. Therefore, signal indicators are utilized for activity/gesture recognition. 

\subsection{\textit{Siamese network}}
The \textit{Siamese network} is a classical architecture generally used in similarity comparison. The major structure of a \textit{Siamese network} is two deep neural networks that share the same weights and architecture. When calculating the similarity between two samples, the inputs of a \textit{Siamese network} are twofold. For example, two different images can be inputted into the \textit{Siamese network} for similarity calculation, two sentences can also be fed into this network to calculate semantic similarity after being transformed as vectors. The outputs of the \textit{Siamese network} are twofold as well because it has two sub-networks. During training, a particular loss named \textit{contrastive loss} is calculated to optimize these two sub-networks. After training, one can establish the similarity of two inputs by comparing the similarity of two sub-networks' outputs. In recent years, the \textit{Siamese network} has also been used for knowledge distillation and model compression.  

\begin{figure*}
    \centering
    \includegraphics[scale=0.78,trim =0 270 80 50,clip]{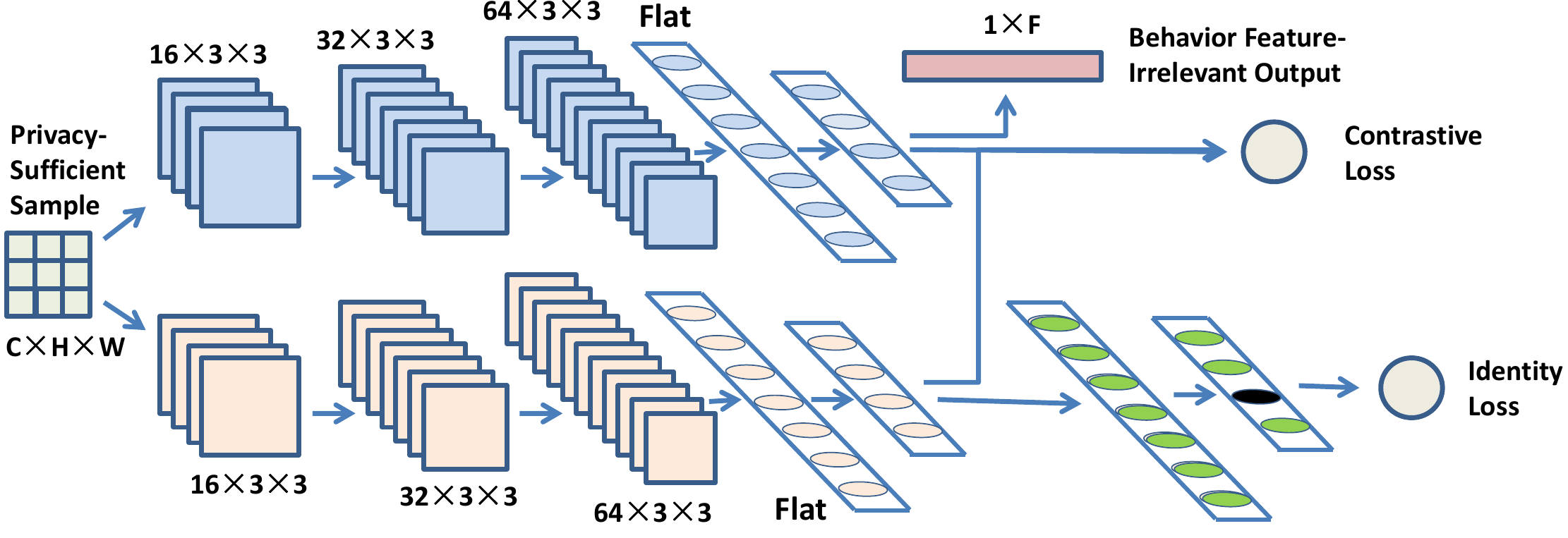}
    \caption{The \textit{Siamese network}-based deep learning model, two kinds of losses: contrastive loss and identity loss are back propagated to optimize the model.}
    \label{fig:siamese}
\end{figure*}

\section{Methodology}
\label{sec:methodology}
We present the framework overview in the first part and build a theoretical model in the second part. The architecture of RFBP-Net is elaborated in the third part. Finally, we introduce the loss functions and the training method of RFBP-Met. 
\subsection{Framework Overview}
In this part, the workflow of our proposed framework is elaborated upon. As shown in Fig. \ref{fig:workflow}, our framework is composed of four modules: \textit{source data validation}, \textit{training set construction}, \textit{processing via RFBP-Net} and \textit{feature quality validation}.

\par \textbf{Source data validation:} This module is the first module of our framework. In this module, we should confirm that the source data is both identity feature-sufficient and behavior feature-sufficient. First, the collected source signal (\textit{i.e.} source data) should be processed so that all the signal samples have the same dimension, e.g. the dimension of $2\times30\times49$ in our RFID experiment. Afterwards, several classic machine/deep learning algorithms/models are selected to classify these signal samples. If the recognition accuracies of identity and behavior are both high, e.g., larger than 80\%, we confirm that the source data is valid and needs to be processed by RFBP-Net. 
    
\par \textbf{Training set construction:} After confirming that the source data is valid, the training set needs to be constructed. This module also contains two steps. In the first step, $2n$ special signal samples are randomly selected. In the second step, these randomly selected signal samples are re-labeled and constructed as $n$ training samples. These two steps are combined together as an algorithm described in a following part.
    
\par \textbf{Processing via RFBP-Net:} This module is the core of our framework and contains three steps. First, since RFBP-Net is a deep model which has special architecture and loss functions, some parameters, e.g., the size of the output feature, need to be set before training. Then, RFBP-Net is trained by using the training set and parameters set in the first step. At last, all $k$ signal samples in the source data are fed into the well-trained RFBP-Net to get $k$ behavior-irrelevant feature vectors.
    
\par \textbf{Feature quality validation:} This module evaluates the quality of the extracted feature, which is extracted from the previous module. This module first selects a suitable algorithm or model to evaluate the feature's quality. Next, due to the fact that our goal is to extract the feature which only contains the identity-relevant feature, it is essential to confirm that the extracted feature can only be used for accurate identity authentication..

\subsection{Theoretical model}
Each signal sample is composed of three components:
\begin{equation}
S = f_M(C_I,C_A,N).
\end{equation}
$f_M(\cdot)$ is the traveling function of the transmitted signals determined by the traveling multi-path in the environment. $C_I$ represents the signal component that contains the identity-relevant feature. $C_A$ represents the signal component that contains the behavior-relevant feature. $N$ is the noise component which should be discarded during feature extraction. 

\par Firstly, the identity-relevant component $C_I$ should be extracted from $S$. The related function,  $f_E(\cdot)$, can be represented by:
\begin{equation}
    C_I = f_E(\Theta_1,S) = f_E(\Theta_1,f_M(C_I,C_A,N)),
\end{equation}
where $\Theta_1$ are the parameters that need to be calculated. Afterwards, we need another function $Re(\cdot)$ to refine $C_I$ so that the output $F_I$, i.e., the identity-relevant feature can be as pure as possible: 
 \begin{equation}
  F_I = Re(\Theta_2,C_I),  
\end{equation}
where $\Theta_2$ are the parameters that need to be calculated as well.

\par In our framework, RFBP-Net realizes both functions of $f_E(\cdot)$ and $Re(\cdot)$. By optimizing the model with \textit{contrastive loss}, the parameters $\Theta_1$ are calculated automatically. Likewise, the parameters $\Theta_2$ are calculated by optimizing the model with \textit{identity loss}.  

\subsection{The architecture of RFBP-Net}
The architecture of RFBP-Net is shown in Fig. \ref{fig:siamese}. The inputs of this model are twofold, i.e., two samples are fed into the model simultaneously. We assume that the RF signal collected to authenticate users also contains behavior privacy. The goal of RFBP-Net is to extract the pure feature that can be used for user authentication while only meagre behavior-relevant feature is contained in the pure feature. 

\par To achieve this goal, RFBP-Net employs a CNN-based deep model as the feature extractor of the \textit{Siamese network}. Specifically, the feature extractor contains three convolutional layers and two fully connected layers. We add a batch normalization function \cite{DBLP:journals/nca/WangMHSZ20} and a ReLU activation function \cite{DBLP:conf/fair2/PretoriusBD19} behind each convolutional layer. The first fully connected layer is followed by a Sigmoid activation function \cite{DBLP:journals/ijiids/Lee14}. In order not to impact the feature representation ability of the feature extractor's output, we do not add any activation function behind the last fully connected layer. The feature extractor is followed by two branches: the first one for \textit{contrastive loss}-based optimization and the second one for \textit{identity loss}-based optimization. The second branch is composed of two fully connected layers and each fully connected layer is followed by a Sigmoid function \cite{DBLP:journals/ijiids/Lee14}.

\begin{algorithm}[t]
\renewcommand{\algorithmicrequire}{\textbf{Input:}}
\renewcommand{\algorithmicensure}{\textbf{Output:}}

  \caption{: Training Set Construction}
  \label{alg:1}
  \begin{algorithmic}[1]
    \REQUIRE Signal sample set with N signal samples: $S = (S_1,S_2,\cdots,S_n)$, corresponding identity label set: $L^\text{SI} = (L^\text{SI}_1,L^\text{SI}_2,\cdots,L^\text{SI}_n)$, corresponding behavior label set: $L^\text{SA} = (L^\text{SA}_1,L^\text{SA}_2,\cdots,L^\text{SA}_n)$, $n \in [1:N]$. 
	\ENSURE Training set: $T = (T_1,T_2,\cdots,T_m)$, corresponding contrastive label set: $L^\text{TC} = (L^\text{TC}_1,L^\text{TC}_2,\cdots,L^\text{TC}_m)$, correspondig identity label set: $L^\text{TI} = (L^\text{TI}_1,L^\text{TI}_2,\cdots,L^\text{TI}_m)$.
	  \STATE $i \leftarrow 0; M \leftarrow 1000; T \leftarrow (\emptyset); L^\text{TC} \leftarrow (\emptyset); L^\text{TI} \leftarrow (\emptyset)$
	  \WHILE{$i < M$}
	    \STATE Selecting signal samples $S_j$ and $S_k$ from $S$ randomly 
	    \IF {$L^\text{SI}_j == L^\text{SI}_k$}
	      \IF{$L^\text{SA}_j \ne L^\text{SA}_k$}
	        \STATE $i \leftarrow i+1$, $T_\text{i} \leftarrow (S_j,S_k)$
	        \STATE Appending $T_\text{i}$, $0$ and $L^\text{SI}_j$ to $T$, $L^\text{TC}$ and $L^\text{TI}$ respectively
	      \ELSE
	        \STATE Discarding $S_j$ and $S_k$
	       \ENDIF
	    \ELSE
	      \IF{$L^\text{SA}_j == L^\text{SA}_k$}
	        \STATE $i \leftarrow i+1$, $T_\text{i} \leftarrow (S_j,S_k)$
	        \STATE Appending $T_\text{i}$, $1$ and $-1$ to $T$, $L^\text{TC}$ and $L^\text{TI}$ respectively
	      \ELSE
	        \STATE Discarding $S_j$ and $S_k$
	       \ENDIF
	    \ENDIF
	  \ENDWHILE
  \end{algorithmic}
  \label{Alg1}
\end{algorithm}

\subsection{Training set}
Training set is crucial for a deep learning model. In our framework, RFBP-Net aims to leverage the knowledge distillation ability of the \textit{Siamese network} to extract specific features. However, The realization of knowledge distillation relies on the basic function of the \textit{Siamese network}: calculating the similarity between two inputs. Hence, we reconstruct the signal samples and re-label them. In detail, because each signal sample has two labels: an identity label and a behavior label, our framework reorganizes the samples by combining two signal sample into one training sample and labeling this training sample based on two rules: 1) A training sample sets `0' as its similarity label if two signal samples in this training sample belong to the same user but different activities. By contrast, the similarity label of the training sample is `1' if the two signal samples of this training sample belong to different users but the same behavior. 2) If the similarity label of the training sample is `0', its second label is set as the corresponding identity label. Otherwise, the second label is set as `-1'.

\par In particular, as described in Alg. 1, we form the training set by randomly sampling from signal samples. One signal sample is randomly selected from all signal samples first, then another signal sample is selected in the same way. Afterwards, those two randomly selected signal samples are organized together to form one training sample based on the aforementioned rules.

\subsection{Loss functions}
Recalling that each training sample contains two different labels: a similarity label and an identity label, two different loss functions are thus utilized to optimize RFBP-Net. Specifically, the \textit{contrastive loss} can be denoted as:
\begin{equation}
  {LOSS}_c = (1-Y_S) (D_W)^2 + Y_S (max\left\{0,{margin}-D_W \right\})^2.  
\end{equation}
In this formula, $Y_S$ is the similarity label, and $D_W$ is the \textit{Euclidean distance} of two inputs which belong to the same training sample. Moreover, $margin$ is empirically set as 3. If we denote $X_1$ and $X_2$ as two inputted signal samples, the \textit{Euclidean distance} $D_W$ can be represented as:
\begin{equation}
  D_W(X_1,X_2) = \sqrt{\left\{G_W(X_1)-G_W(X_2)\right\}^2}. 
\end{equation}
\par In order to improve the ability of the identity-relevant feature extraction of RFBP-Net, we introduce the identity loss into the optimization step. In detail, RFBP-Net utilizes \textit{cross entropy loss} \cite{DBLP:journals/corr/abs-1901-08360} to calculate the identity loss. The loss function can be denoted by:
\begin{equation}
  {LOSS}_P = -\sum_\text{c=1}^M y_c \log(P_c),  
\end{equation}
in which $y_c$ is the indication variable, $P_c$ is the probability that targeting sample belongs to class $c$ and $M$ is the number of classes.

\subsection{Objective and training}
Ultimately, the final loss of the optimization objective can be represented by: \begin{equation}
  {LOSS}_F = \alpha {LOSS}_C + (1-\alpha) {LOSS}_P ,\quad \alpha \in [0,1].  
\end{equation}
During training, $n$ training samples are divided into $k$ batches and fed into the model. The number of training periods is set as $p$. Empirically, as default, $n$, $k$ and $p$ are respectively set as 1000, 10 and 200. In order to fit the special requirements of the performance in some special scenarios (\textit{e.g.}, a high-level protection of RFBP is in demand yet the requirement for user recognition accuracy is not acute), a ratio trade-off between the identity-relevant feature and the behavior-relevant feature can be adjusted by altering $\alpha$ based on the requirement of the specific application scenario.

\section{Evaluation with RFID}
\label{sec:experiment}
In order to evaluate the performance of our framework with RFID signals, we conducted experiments with five volunteers and collected over 4000 signal samples. The ages the of volunteers varied from 21 to 31 and the heights of them varied from 165 to 188 centimeters (2 females and 3 males).  

\par \textbf{Hardware: }The reader used for signal modulation and demodulation was a COTs reader whose type was Impiji R420. It was connected with a commercial one-dimensional Larid A9028 antenna. We build a tag array with size $7 \times 7$. The type of the tags was Alien-9629.
\par \textbf{Software: }We used \textit{Visual Studio} and \textit{C\#} to control the transmission and receiving procedure of the RFID system. To avoid transmission collision, we employed standard frame-based slot-ALOHA protocol to arrange the response time of each tag. The signal processing was completed by using MATLAB. The feature extraction model was built through the standard deep learning framework \textit{Pytorch} and hence the code was programmed in \textit{Python} language in \textit{Eclipse}. Likewise, the \textit{feature quality evaluation} was achieved by using Python language as well.

\begin{figure}[t] 
\centering    
\subfigure[Experiment setup with RFID system.] {
\includegraphics[scale=0.27, trim = 0 20 280 0,clip]{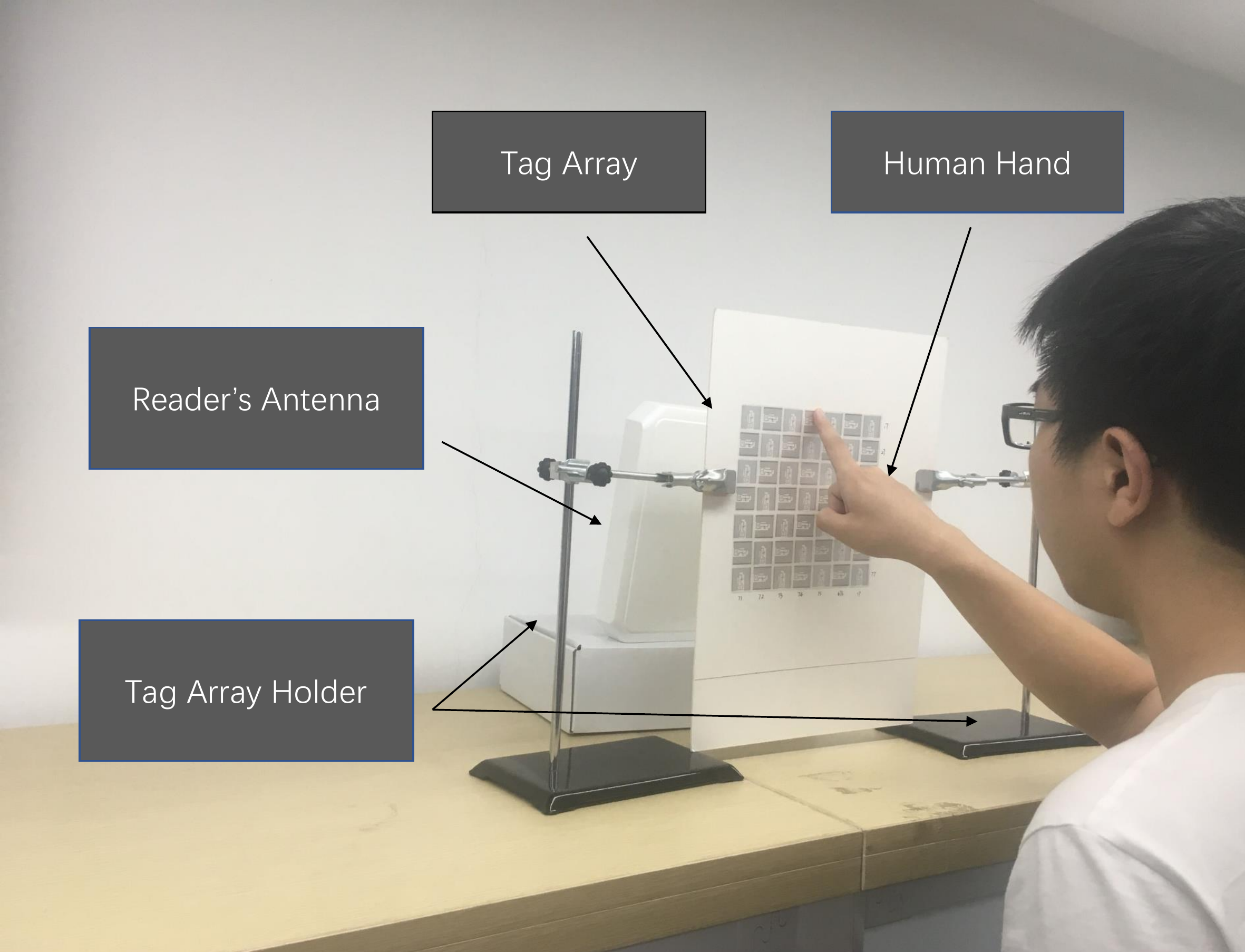}  
}
\centering
\subfigure[Ten writing numbers from 0 to 9.] { 
\includegraphics[scale=0.75,trim=0 425 440 0,clip]{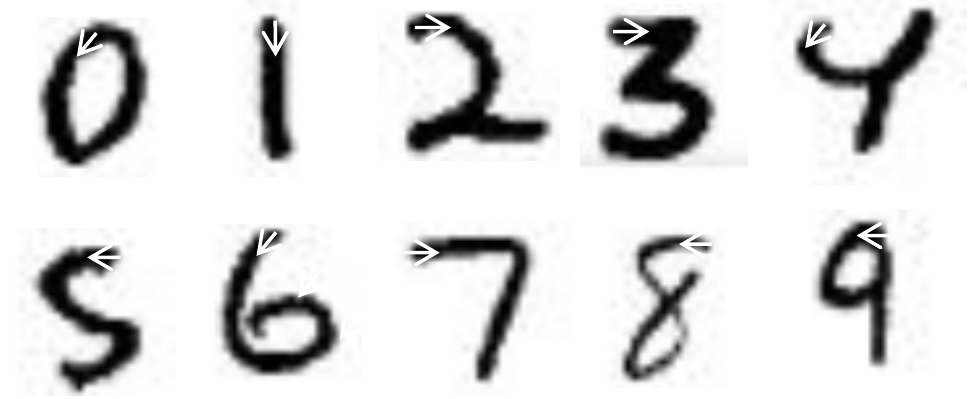}
}
\caption{Experiment setup.}     
\label{fig:experiment_setup}     
\end{figure}

\par \textbf{Experiment setup: }As shown in Fig. \ref{fig:experiment_setup}(a), we employed a commercial RFID system for the transmission and receiving of RF signals. Volunteers were asked to write ten numbers, as shown in Fig. \ref{fig:experiment_setup}(b), to represent ten activities. The white arrow on the number is the start point of writing and the direction of the arrow is the writing direction. 

\par \textbf{Data preprocessing: }The received signal samples, which were time-series data originally, need to be processed to have a regular shape. In our experiments, RSS values and phase values are first formed as value array and then 30 value arrays are piled together to form a 3-dimensional signal sample which has the  dimension of $2\times 30 \times 49$. 

\subsection{Validity of source data}
In this part, the signal samples that need to be processed by RFBP-Net are termed as \textit{source data}. Before evaluating the performance of our framework, the validity of \textit{source data} needs to be verified, i.e., it is vital to prove that sufficient identity-relevant features and activity-relevant features are contained in the \textit{source data}. We first trained five learning models: k-nearest neighbours (KNN), naive Bayes (NB), support vector machine (SVM), normal neural network with two fully-connected layers (NN) and convolutional neural network (CNN). Then we tested them with \textit{source data}. Specially, when training the first five learning models, we normalized the \textit{source data} via min-max-normalization, which can be represented by:
\begin{equation}
  X_{nor} = \dfrac{X_{ori}-X_{min}}{X_{max}-X{min}}.  
\end{equation}
In this formula, the original attribute value and normalized attribute value are denoted as $X_{ori}$ and $X_{nor}$ respectively. $X_{min}$ and $X_{max}$ mean the minimal attribute value and the maximal attribute value in all training samples. The experiment results are shown in Table \ref{table1}, from which one can find that by using KNN, NN or CNN, the authentication accuracy of user identity is larger than 98.90\%. By contrast, NB, SVM, and DT perform relatively worse. As for the accuracy of activity recognition, both CNN and KNN achieve 94.74\%, yet CNN is more compatible for constructing a deep model i.e., the \textit{Siamese network}. Based on the above experiment results, two conclusions can be reached, summarized as: 1) It is apparent that sufficient identity-relevant features and activity-relevant features can be provided by \textit{source data} for accurate identity authentication and activity recognition. 2) CNN is qualified to be employed as the fundamental architecture of our deep model.

\begin{table*}
\begin{center}
{\caption{Confirming that training set is sufficient in both kinds of features.}\label{table1}}
\begin{tabular}{ccccccc}

\cline{2-7}
\rule{0pt}{12pt}
Goal&K-Nearest Neighbours&Naive Bayes&Support Vector Machine&Decision Tree&Neural Network&CNN\\
\hline
\\[-6pt]
\quad Identity&99.50\%&63.89\%&93.35\%&62.60\%&100.00\%&99.95\%\\
\quad Activity&95.63\%&27.38\%&94.64\%&48.12\%&89.78\%&94.74\%\\
\hline

\end{tabular}
\end{center}
\end{table*}

\begin{table*}
\begin{center}
{\caption{The learning model selection for feature quality evaluation.}\label{table2}}
\begin{tabular}{ccccccc}
\cline{2-6}
\rule{0pt}{12pt}
Recognition Goal&K-Nearest Neighbours&Naive Bayes&Support Vector Machine&Decision Tree&Neural Nnetwork\\
\hline
\\[-6pt]
\quad Identity&91.87\%&90.08\%&63.19\%&27.38\%&95.63\%\\
\quad Activity&17.86\%&13.29\%&23.12\%&17.16\%&23.90\%\\
\hline

\end{tabular}
\end{center}
\end{table*}

\begin{figure*}[t] 
\centering    
\subfigure[The effect of training set size.] { 
\includegraphics[scale=0.31,trim=5 0 50 20,clip]{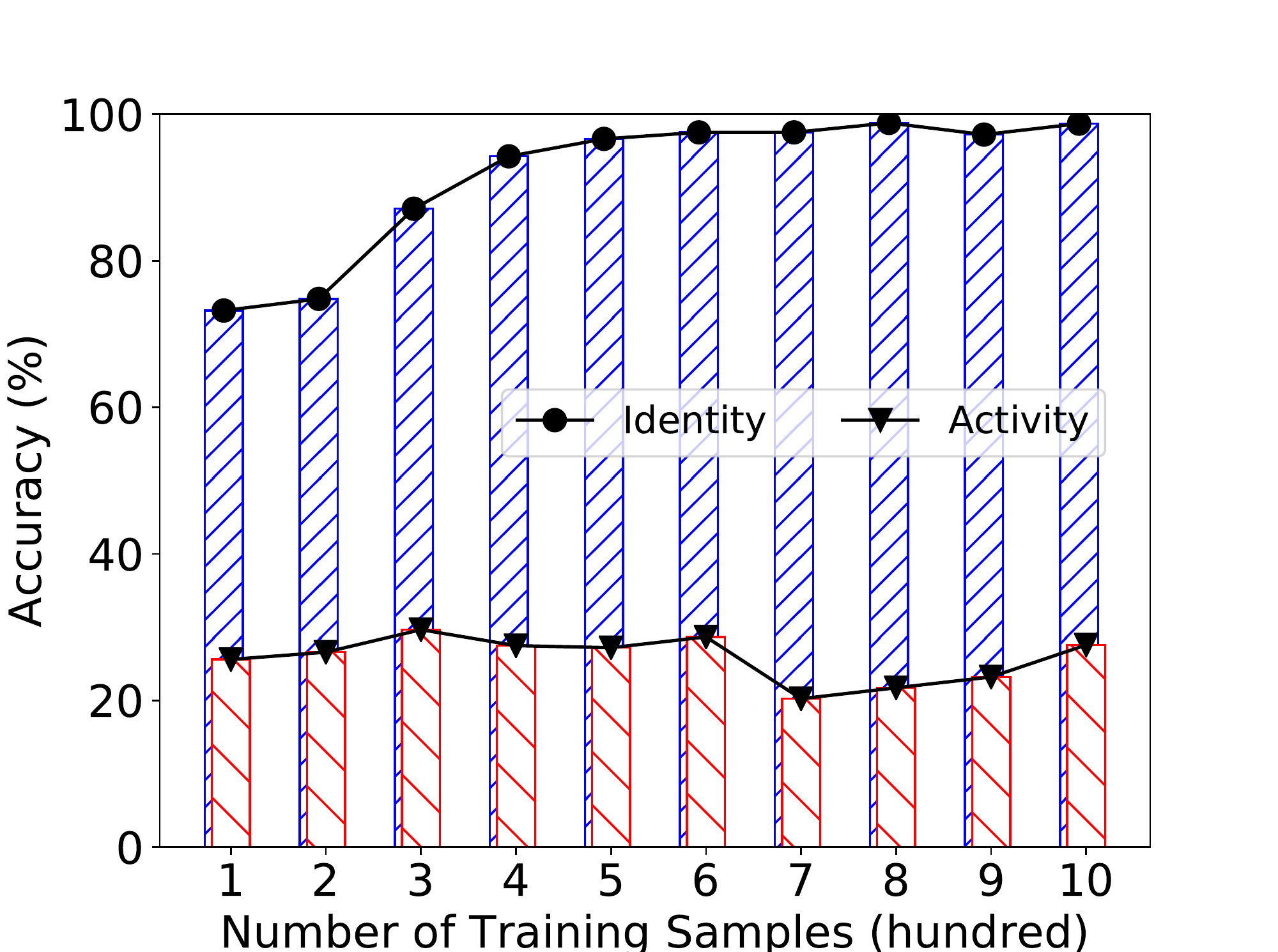}
}
\subfigure[The effect of $\alpha$.] { 
\includegraphics[scale=0.3,trim=5 0 50 20,clip]{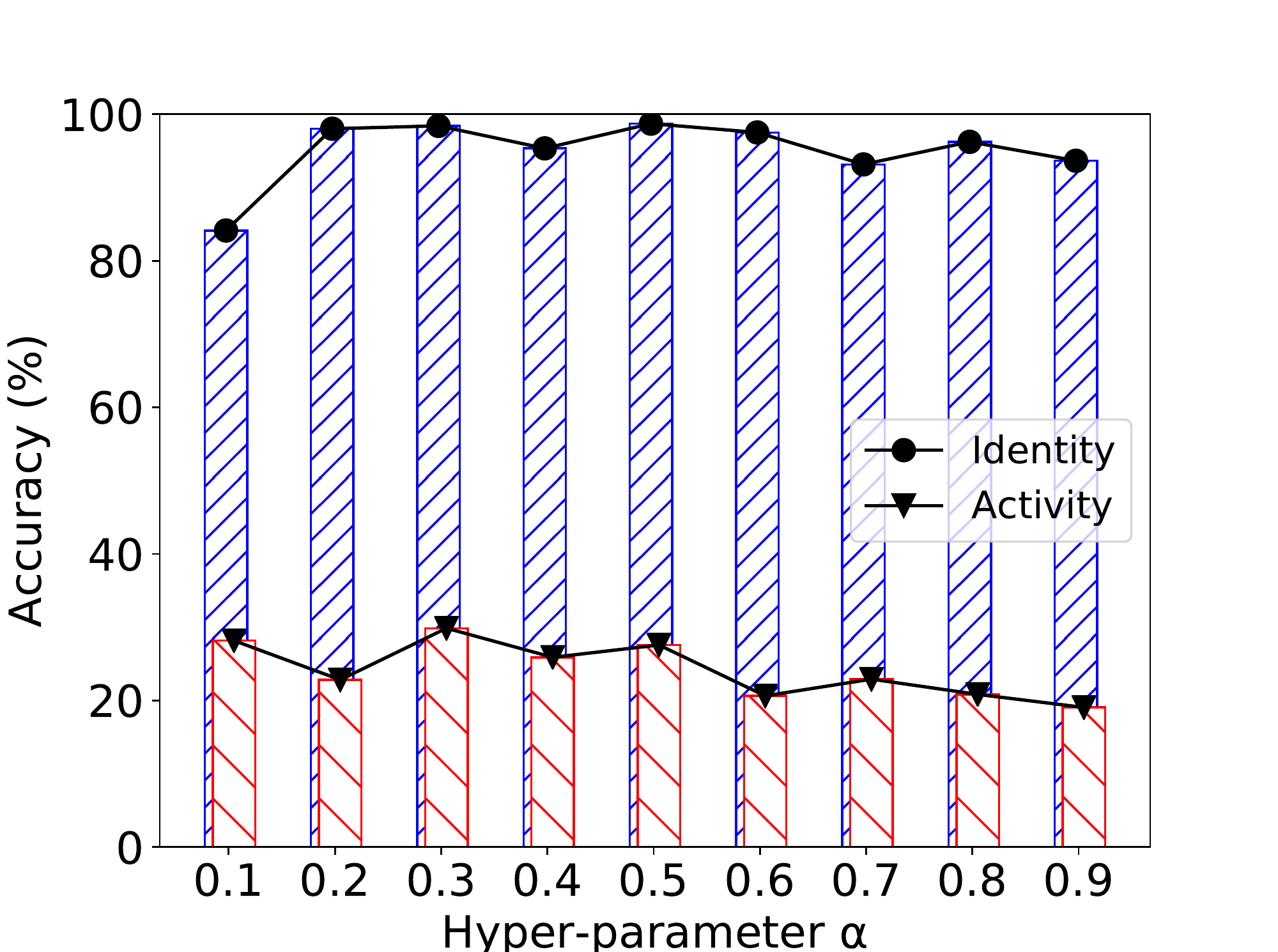}
}
\subfigure[The effect of the feature size.] { 
\includegraphics[scale=0.31,trim=5 0 50 20,clip]{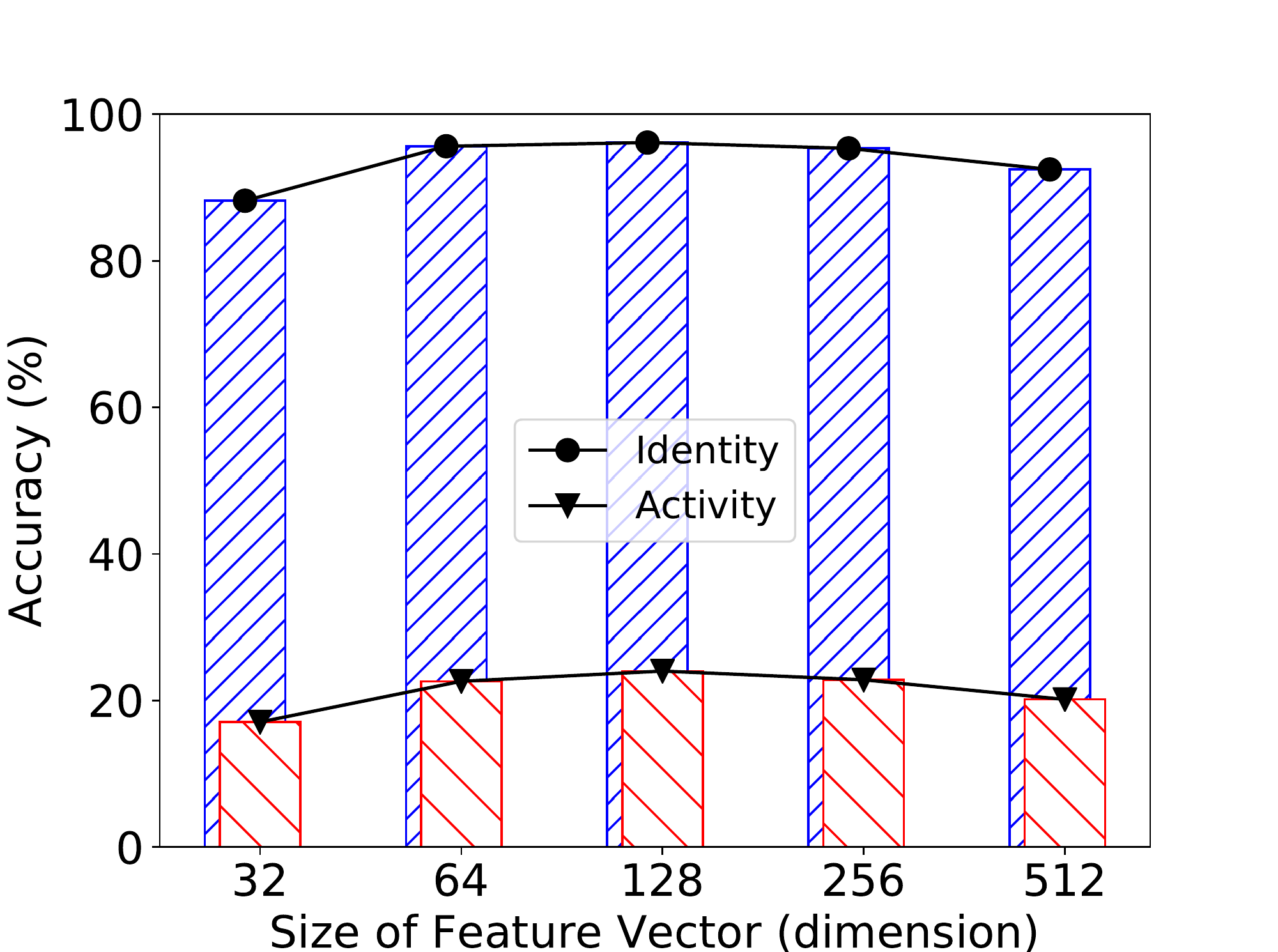}
}
\caption{The effects of (a) training set size, (b) $\alpha$ and (c) feature size.}     
\label{fig:training_alpha_feature}     
\end{figure*}

\par Therefore, those two conclusions confirm the rationality of the design of our \textit{Siamese network}-based deep model.

\subsection{Learning model selection for feature quality evaluation }
Only using one learning model is insufficient to demonstrate that extracted features are activity-privacy-irrelevant while yielding high authentication accuracies of user identity. Therefore, as shown in Table \ref{table2}, we use five learning models to evaluate the feature quality. The default hyper-parameters $\alpha$, feature size (\textit{i.e.}, the number of the elements in the extracted feature vector), and size of training set (\textit{i.e.}, the number of the training samples) are 0.5, 64 and 1000 respectively. The reason that CNN is discarded is that the shape of the extracted feature vector is a 1-dimensional vector, which is not compatible for 2-dimensional convolution. Table \ref{table2} shows that KNN and NN retain an excellent performance on identity authentication compared with other learning models. Though the recognition accuracy of identity drops almost 5\% by using NN, the reduction of the activity recognition accuracy is far larger than 5\%, i.e., the accuracy of activity recognition drops over 70\%. Ultimately, two conclusions can be reached according the experiment results: 1) The proposed deep model is qualified to extract high-quality features that can be utilized for accurate user authentication while protecting activity privacy. 2) It is reasonable to employ NN for feature quality evaluation.

Thus the criteria for judging whether the extracted feature vector is high-quality is: the higher the accuracy of identity authentication, and the lower the accuracy of activity recognition, the higher quality the extracted features . Moreover, NN is used to evaluate the feature quality in all the remaining extensive experiments due to its outstanding recognition capability compared with the other four learning models. 

\subsection{Effect of training set size}
The volume (\textit{i.e.}, size) of the training set is one of the significant factors that directly influences the quality of the extracted features. It is worth noting that \textit{Siamese network} is qualified to perform well even with a small training set. In order to guarantee that the only variable is the size of the training set. we first randomly selected 1000 training samples as the fundamental training set and conducted all the experiments with the subsets of the fundamental training set. 
\par We varied the size of the training sets from 100 to 1000. The experiment results are shown in Fig. \ref{fig:training_alpha_feature}(a). The results demonstrate that when the value of the horizontal axis is smaller than 7, with the increase of the size of the training set, the identity authentication accuracy keeps increasing and the activity recognition accuracy remains relatively stable (around 25\%). Furthermore, when the size of the training set continues increasing, the accuracy of activity recognition starts decreasing (deceasing to 20\% approximately). Afterwards, the accuracy of activity recognition rebounds to 25\% approximately, while the accuracy of identity authentication achieves 98\%. 

\par Though both the identity authentication accuracy curve and the activity recognition accuracy curve become flat when the training set size is larger than 600, in order to guarantee that the extensive experiments were not influenced by the training set size and the training samples were sufficient for model training, we fixed the size of the training set as 1000 in the following experiments.  

\begin{figure}
    \centering
    \includegraphics[scale=0.55,trim=5 0 50 10,clip]{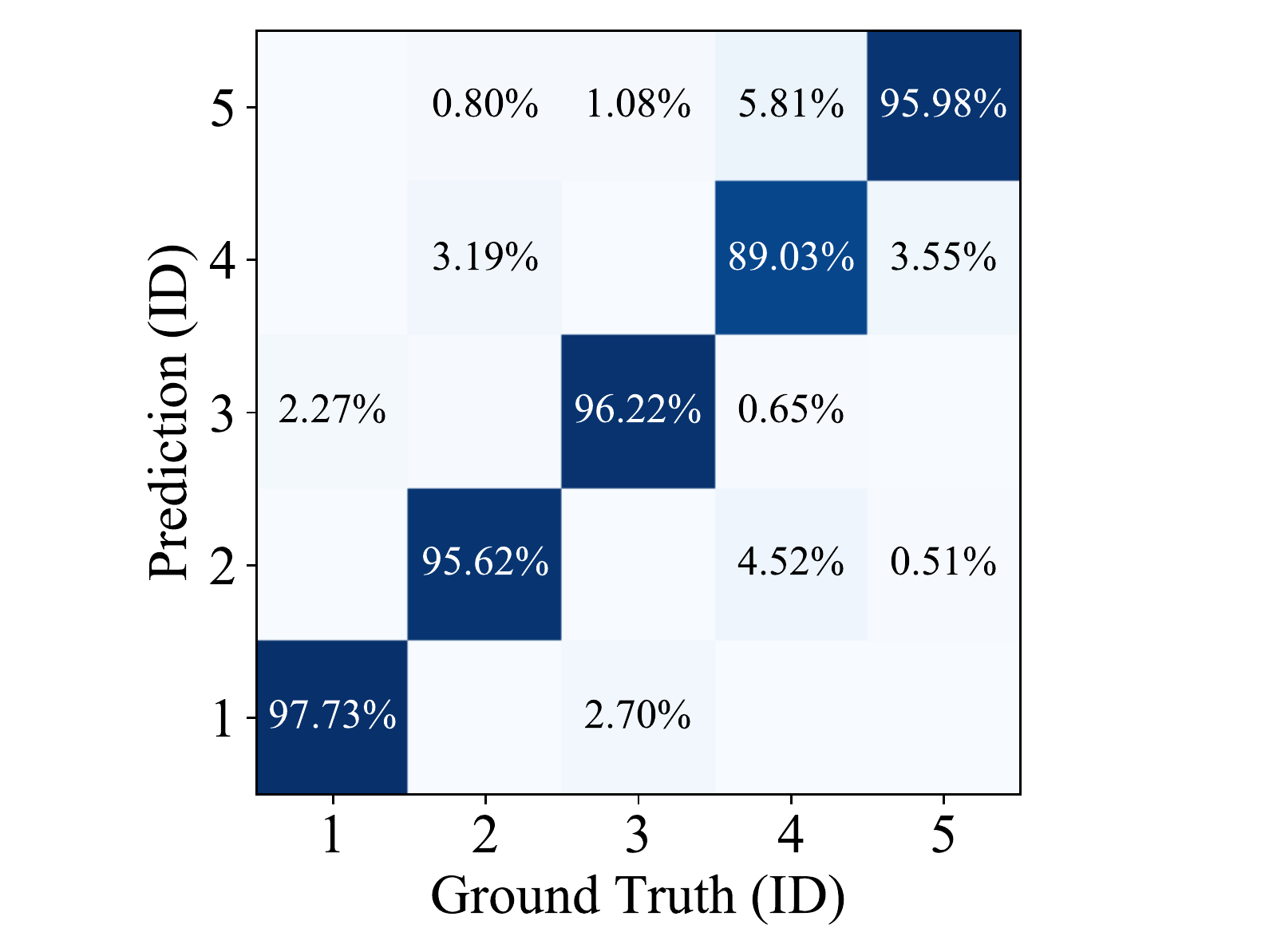}
    \caption{The confusion matrix of identity authentication.}
    \label{fig:id_CM}
\end{figure}

\subsection{Effect of the hyper-parameter}
In this experiment, we hypothesized that the hyper-parameter $\alpha$ was related to the trade-off between the recognition accuracy of activity and the authentication accuracy of identity. In order to validate our hypothesis, we varied $\alpha$ from 0.1 to 0.9 with a stride of 0.1. The experiment results in Fig. \ref{fig:training_alpha_feature}(b) show that when $\alpha$ approximates 0.5, the identity authentication accuracy is significantly high and the activity recognition accuracy is low. Yet, when $\alpha$ is too small or too large, the feature quality does not meet our expectation, e.g., still high identity authentication accuracy and low activity recognition accuracy with no trade-off in between. 


\subsection{Effect of feature size}
In order to explore whether the change of the extracted feature size (the number of elements in the feature vector) affects the performance of RFBP-Net, we respectively set the feature size as 32, 64, 128, 256 and 512 to estimate the feature quality. The experiment result shown in Fig. \ref{fig:training_alpha_feature}(c) demonstrates that with the increase of the feature size, both kinds of accuracies first increase and then decrease. Two curves reach respective peaks when the size of the feature vector is 128. This regularity of variation coincides with the variation regularity in Fig. \ref{fig:training_alpha_feature}(c). Besides, though the identity authentication accuracy reaches a maximum (96.13\%) when the feature size is 128, the activity recognition achieves its peak (24.01\%) as well. Thus, 128 is not the best choice for feature extraction. It can be noticed that at 64 (horizontal axis), the accuracies of identity authentication and activity recognition are 95.63\% and 22.62\% respectively, where the absolute value of the accuracy difference reaches the maximum. Therefore, it is reasonable that we set 64 as the default of feature size.

\begin{table}
\begin{center}
{\caption{Comparing RFBP-Net with random guess.}\label{tab:comparison}}
\begin{tabular}{ccccccc}
\cline{2-4}
\rule{0pt}{12pt}
Status&Ran. Gue.&Original Dataset&RFBP-Net\\
\hline
\\[-6pt]
\quad Identity&20\%&100\%&95\%\\
\quad Activity&10\%&95\%&25\%\\
\hline
\end{tabular}
\end{center}
\end{table}

\subsection{Evaluation with well-selected parameters}
Following the evaluation results above, we set the size of the training set, the hyper-parameter $\alpha$ and the size of the extracted feature vector as 1000, 0.7 and 64 respectively because RFBP-Net could achieve the best performance under this condition in this experiment. The confusion matrix of identity authentication is shown in Fig. \ref{fig:id_CM} , where one can see that the identity-relevant feature is effectively retained because the colors on the diagonal are significantly deeper than surrounding colors. Meanwhile, the activity-relevant feature is effectively reduced because the activity recognition accuracy drops more than 72\%..

\subsection{Comparison with random guess}
Since we are the first to propose the concept of RFBP and also the first to solve the RFBP preserving issue in AOD, there is no related work that can be used for comparison. However, we can compare the authentication accuracy and recognition accuracy of RFBP-Net with a random guess to show the superiority of RFBP-Net. The comparison results are shown in Table \ref{tab:comparison}. The results demonstrate that the privacy-preserved dataset, i.e., the dataset processed by RFBP-Net, provides equally identity authentication accuracy as the original dataset and as low activity recognition accuracy as a random guess. Thus, RFBP-Net is effective in activity privacy preserving. 

\begin{figure}[t] 
\setlength{\belowcaptionskip}{-0.5cm}
\centering    
\subfigure[Experiment setup with WiFi system.] {
\includegraphics[scale=0.56, trim = 10 300 300 30,clip]{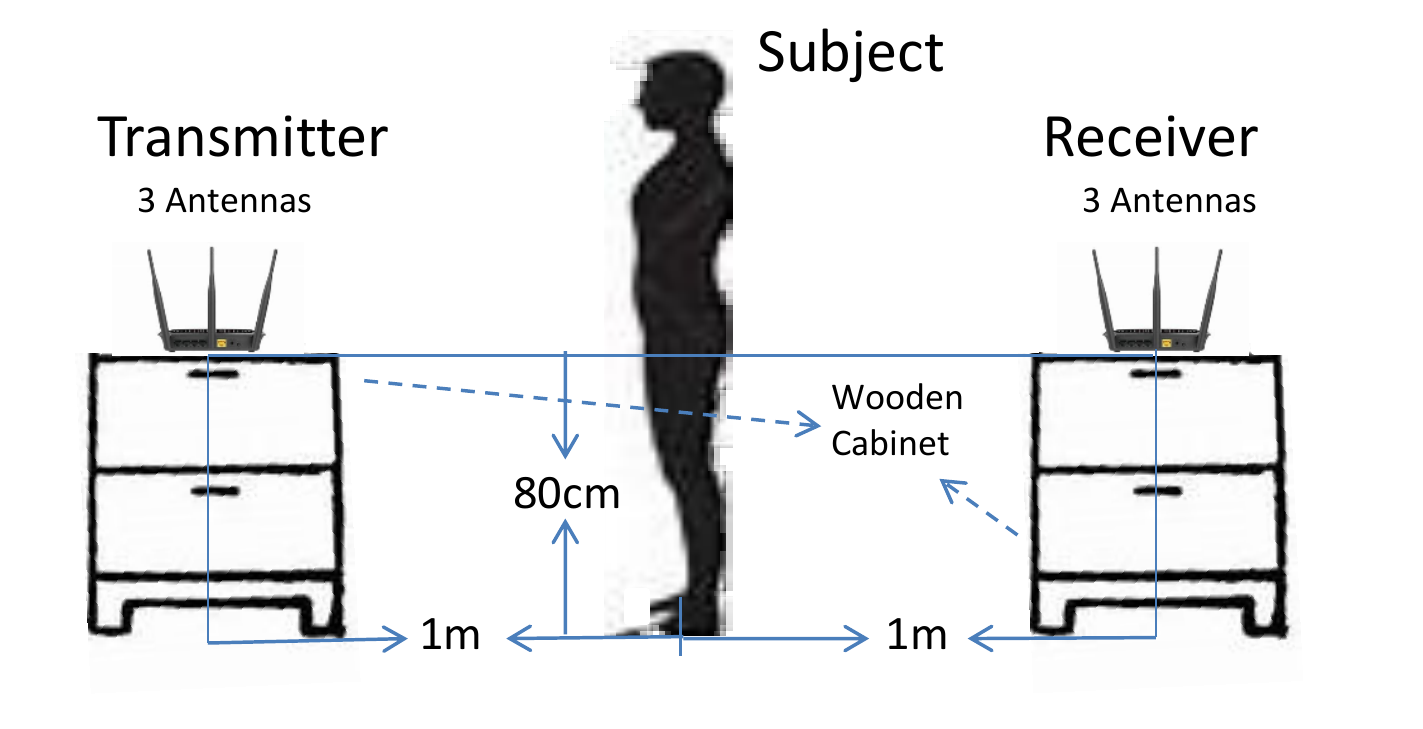}  
}
\centering
\subfigure[Ten gesture numbers from 0 to 9.] { 
\includegraphics[scale=0.4,trim=0 250 200 0,clip]{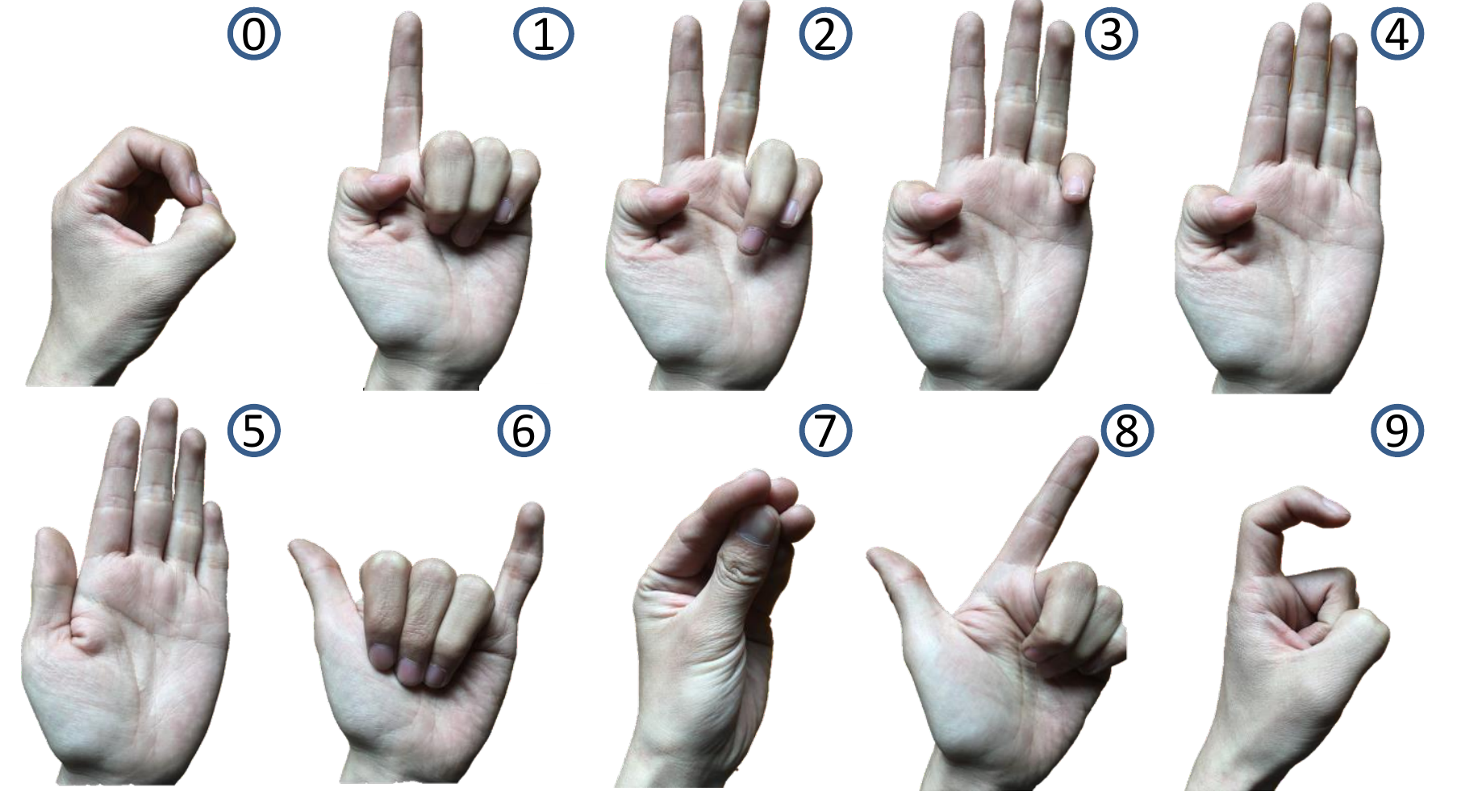}
}
\caption{Experiment setup.}     
\label{fig:experiment_setup_wifi}     
\end{figure}

\begin{table*}
\begin{center}
{\caption{Source data is sufficient in both kinds of features.}\label{feature_sufficient_wifi}}
\begin{tabular}{ccccccccccccc}
\cline{2-13}
\rule{0pt}{12pt}
Option&Sub.1&Sub.2&Sub.3&Sub.4&Sub.5&Sub.6&Sub.7&Sub.8&Sub.9&Sub.10&Avg. Acc.&ID\\
\hline
\\[-6pt]
\quad KNN&73.69\%&78.07\%&72.26\%&63.57\%&65.14\%&82.19\%&83.19\%&70.30\%&72.79\%&77.99\%&73.86\%&99.63\%\\
\quad NB&75.92\%&78.28\%&72.50\%&70.15\%&68.01\%&83.74\%&67.83\%&71.30\%&68.04\%&82.63\%&73.92\%&98.39\%\\
\quad SVM&87.68\%&84.17\%&84.93\%&57.95\%&79.61\%&90.73\%&86.63\%&83.75\%&78.99\%&83.74\%&83.80\%&99.56\%\\
\quad DT&75.48\%&72.46\%&72.83\%&77.73\%&72.40\%&79.78\%&73.33\%&68.25\%&79.59\%&83.53\%&75.54\%&97.83\%\\
\quad NN&96.58\%&95.25\%&96.18\%&84.45\%&85.47\%&97.82\%&95.97\%&95.01\%&95.36\%&94.81\%&93.69\%&99.79\%\\
\quad CNN&99.73\%&99.47\%&99.73\%&99.06\%&99.47\%&99.20\%&99.07\%&99.20\%&98.03\%&99.53\%&99.15\%&99.70\%\\
\hline
\end{tabular}
\end{center}
\end{table*}

\begin{table*}
\begin{center}
{\caption{The learning model selection for feature quality evaluation.}\label{classifier_selection_wifi}}
\begin{tabular}{ccccccccccccc}
\cline{2-13}
\rule{0pt}{12pt}
Option&Sub.1&Sub.2&Sub.3&Sub.4&Sub.5&Sub.6&Sub.7&Sub.8&Sub.9&Sub.10&Avg. Acc.&ID\\
\hline
\\[-6pt]
\quad KNN&11.01\%&18.80\%&14.39\%&14.80\%&10.37\%&25.58\%&20.23\%&19.08\%&15.95\%&18.25\%&16.85\%&99.70\%\\
\quad NB&14.19\%&18.17\%&18.66\%&17.31\%&15.02\%&23.49\%&17.53\%&20.27\%&18.96\%&21.83\%&18.54\%&98.49\%\\
\quad SVM&11.72\%&20.89\%&17.24\%&17.15\%&12.49\%&27.01\%&19.05\%&21.84\%&17.34\%&20.70\%&18.51\%&99.73\%\\
\quad DT&12.30\%&16.07\%&14.24\%&15.15\%&10.56\%&21.61\%&17.80\%&16.73\%&13.87\%&16.29\%&17.14\%&99.67\%\\
\quad NN&16.32\%&21.84\%&15.49\%&21.72\%&16.70\%&21.01\%&17.96\%&18.40\%&16.36\%&25.72\%&19.15\%&99.51\%\\
\hline
\end{tabular}
\end{center}
\end{table*}

\section{Evaluation with WiFi}
\label{experiment_wifi}
In order to evaluate the performance of RFBP-Net with WiFi signal, we conducted experiments with ten volunteers and collected over 29000 signal samples. The ages of volunteers varied from 22 to 35 and the their heights varied from 160 to 188 centimeters (2 females and 8 males).

\par \textbf{Hardware: }we used a transmitter which had three antennas to emit WiFi signals. A router which also has three antennas was employed as the receiver. The router was made by TPLink and the type was WDR7500-V3. Each transceiver was equipped with an Atheros Chip whose type was AR9500. 

\par \textbf{Software: }we use \textit{Linux} operation system to collect WiFi signals. We used an off-the-shelf Linux CSI tool \cite{DBLP:journals/ccr/HalperinHSW11} to measure the CSI of WiFi signals. The transmission rate was 100 packets per second and we used 56 sub-channels. The collected raw data was first processed by \textit{MATLAB} using Butterworth filter. Then the filtered signal was segmented via \textit{Python} and \textit{ECLIPSE}. Finally, the architecture of RFBP-Net was coded by \textit{Pytorch}. The model was trained also by using \textit{ECLIPSE}.

\par \textbf{Experiment setup: }as shown in Fig. \ref{fig:experiment_setup_wifi}(a), the transmitter was placed two meters away from the receiver. While posing the gestures shown in Fig. \ref{fig:experiment_setup_wifi}(b), the volunteer was standing in between. Both the transmitter and the receiver were placed on wooden cabinets, whose top surfaces were 80 centimeters off the ground. In this way, the main path of the WiFi signal could approximately pass through the volunteer's hand.

\par \textbf{Data preprocessing: }after raw WiFi signal collection, we filtered raw data with a 5th-order low pass Butterworth filter with a cutoff frequency of 0.1HZ. Afterwards, we segmented the time-series data of each gesture of each volunteer so that each signal sample had the dimension $504\times10$. The first dimension 504 is 56 sub-channels $\times$ 3 transmission antennas $\times$ 3 receiving antennas. The second dimension 10 is 10 time stamps.

\subsection{Validity of source data}
As shown in Table \ref{feature_sufficient_wifi}, we separately trained KNN, NB, SVM, DT, NN and CNN by using 75\% source signal samples and tested with 25\% source signal samples. The columns from Sub.1 to Sub.10 mean the gesture recognition accuracy of ten different volunteers. Avg. Acc. means the average gesture recognition accuracy of these ten volunteers. ID means the identity authentication accuracy. Since the gesture recognition accuracy of WiFi is highly related to the domain of experiment components \cite{DBLP:conf/mobisys/ZhengZ0ZLW019}, we classified the gestures of each subject (volunteer) one by one rather than all the subjects together. As can be seen from the results, all the average accuracies of gesture recognition are higher than 73\% and all the identity authentication accuracies are higher than 97\%. In NN and CNN, both kinds of accuracy are even higher than 93\%. These results demonstrate that the source data is both identity feature-sufficient and gesture feature-sufficient. Moreover, CNN is an outstanding choice as the basic architecture of \textit{Siamese network}. Thus in the following evaluation part, we use CNN to construct the \textit{Siamese network}.

\begin{figure}[t]
    \centering
    \includegraphics[scale=0.4,trim = 0 0 10 10,clip]{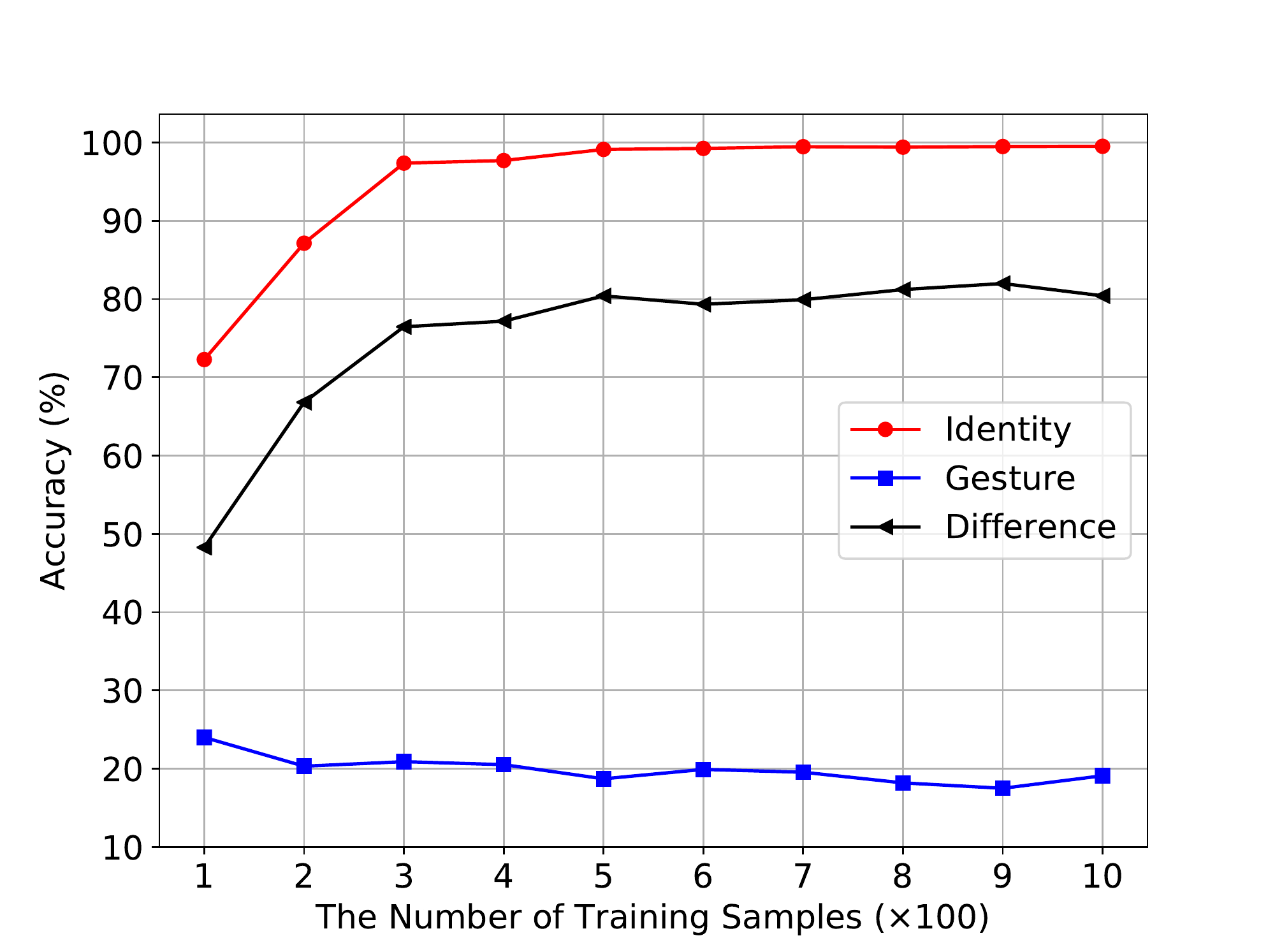}
    \caption{The effect of training set size.}
    \label{fig:training_set_wifi}
\end{figure}

\begin{figure*}[t] 
\centering    
\subfigure[The effect of the hyper-parameter $\alpha$.] {
\includegraphics[scale=0.32,trim=20 2 55 40,clip]{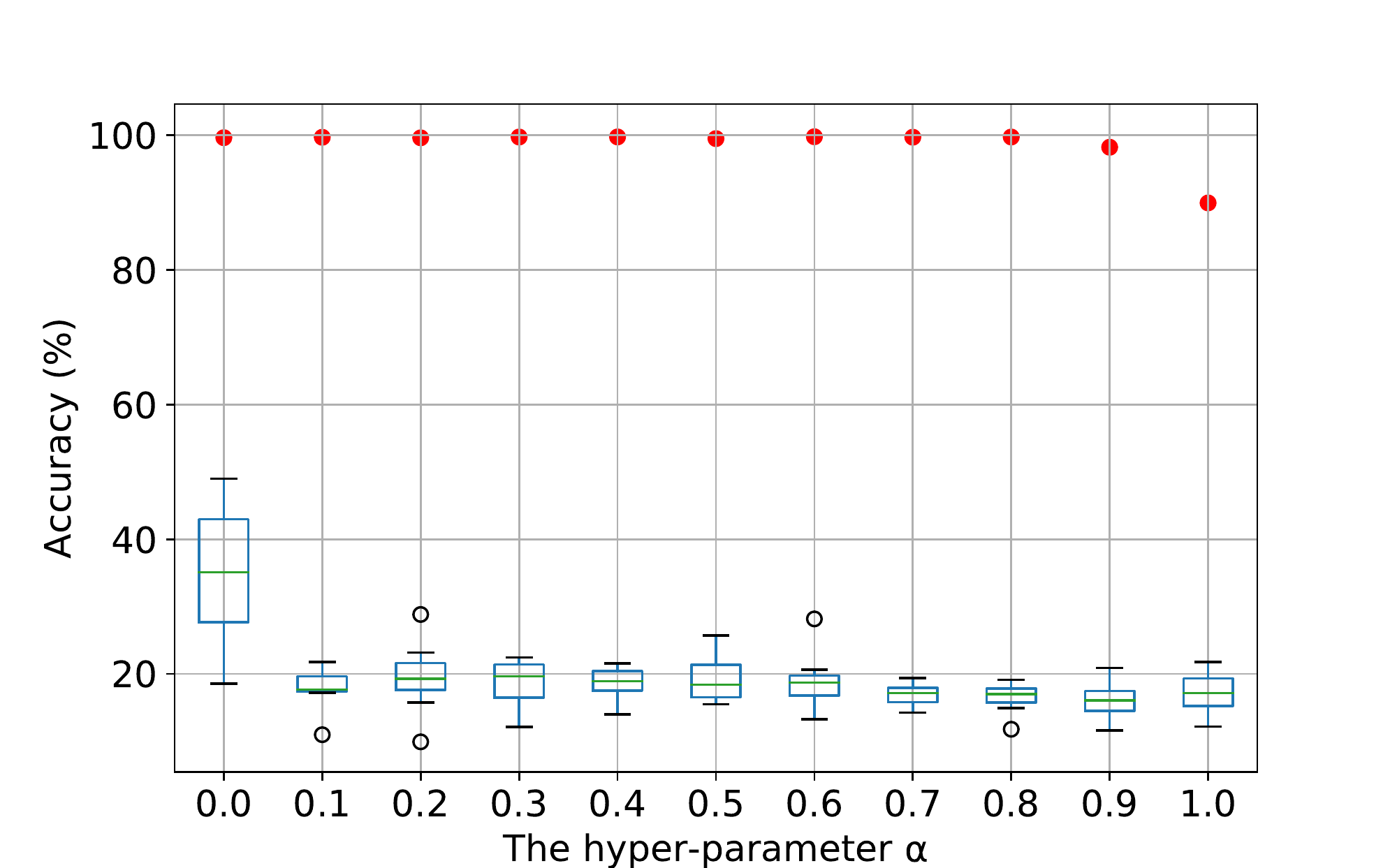}  
}
\subfigure[The effect of the feature size.] { 
\includegraphics[scale=0.32,trim=20 2 55 40,clip]{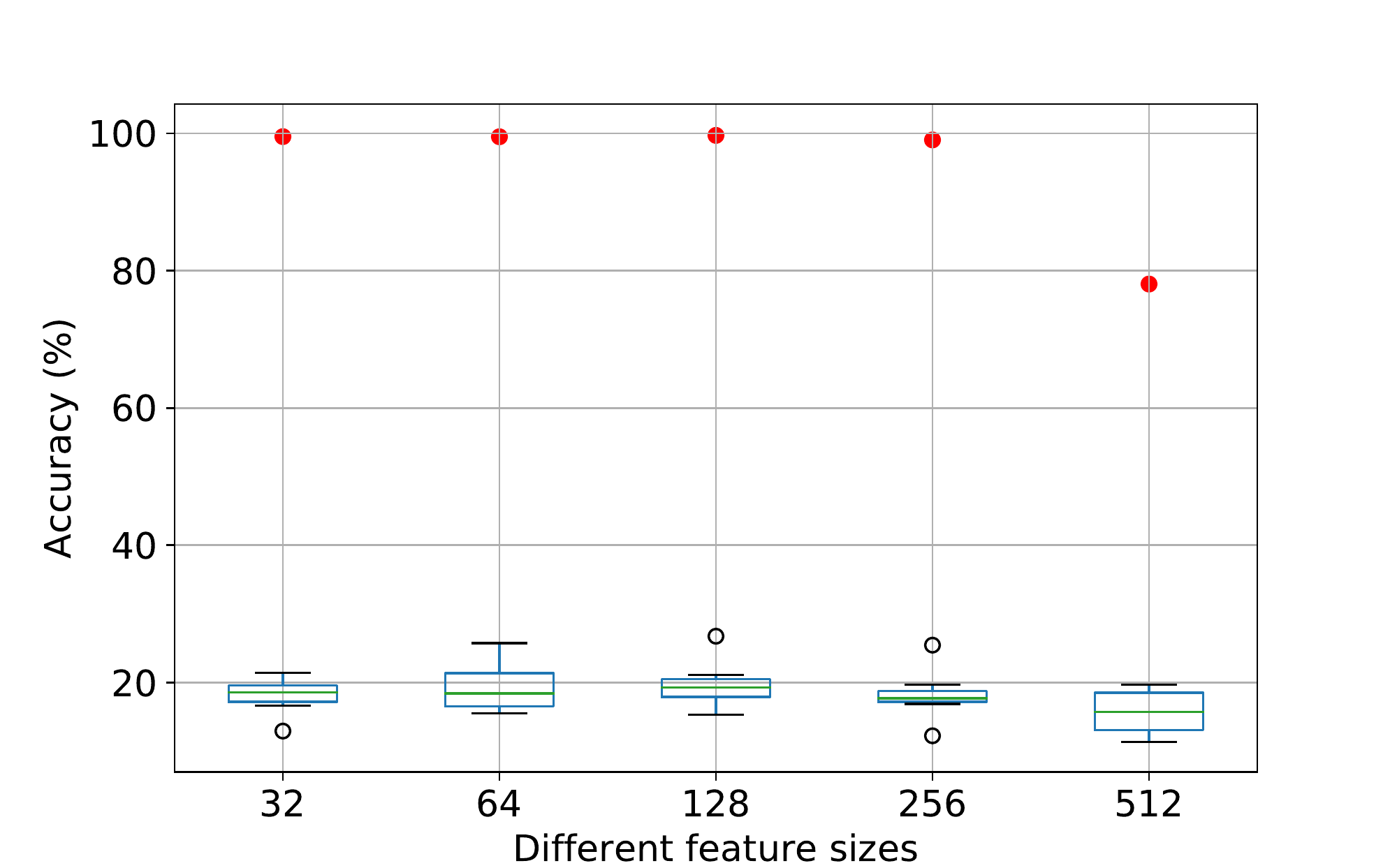}
}
\subfigure[The effect of activation functions.] { 
\includegraphics[scale=0.32,trim=17 5 50 40,clip]{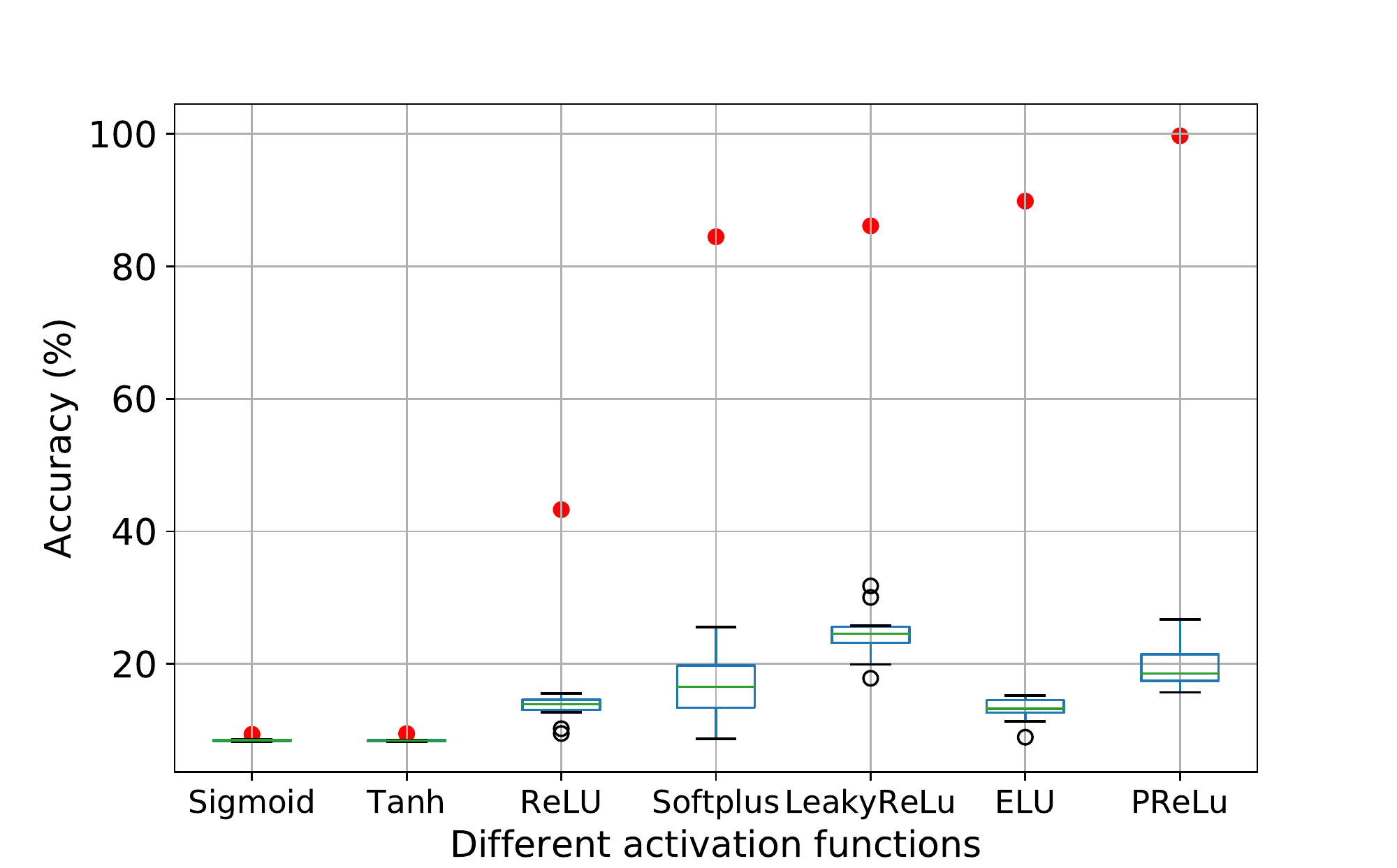}
}
\caption{The effects of (a) $\alpha$, (b) feature.size and (c) activation functions. Red points represent identity authentication, the blue box represents the gesture recognition.}     
\label{fig:alpha_size_activation}  
\end{figure*}

\subsection{Learning model selection for feature quality evaluation}
After training set construction, we first set the training set size, $\alpha$ and feature size as 1000, 0.5 and 64, respectively. Then we trained KNN, NB, SVM, DT and NN and displayed the results in Table \ref{classifier_selection_wifi}. From the column of Avg. Acc. one can find that by using the extracted feature, the average gesture recognition accuracies of all learning models have dropped lower than 20\%. Meanwhile, the identity authentication accuracies of all learning models are higher than 98\%. To our surprise, the identity authentication accuracy of DT even increases by 1.84\%. Since NN shows the highest gesture recognition, we use NN to evaluate the feature quality in the following parts.

\subsection{Effect of training set size}
In a similar way to the RFID evaluation method, we set the number of training samples from 100 to 1000 with a stride of 100 to explore the effect of the training set size. The randomness of the training set would introduce extra variables, which would blur the accuracy variation caused by the training set size. Thus, we only constructed the training set once and use it in this and all following experiments. The experiment results are shown in Fig. \ref{fig:training_set_wifi}. The top curve, middle curve and bottom curve represent identity authentication accuracy, accuracy difference and gesture recognition accuracy, respectively. When the number of training samples is smaller than 500, three curves have positive gradients. After 500, these three curve become flat. Thus, 500 training samples are sufficient for WiFi-based RFBP-Net training. In order to guarantee that RFBP-Net is training by using sufficient training samples, we set the training set size as 1000 in the following experiments.

\subsection{Effect of the hyper-parameter}
We varied $\alpha$ from 0.0 to 1.0 with the a of 0.1 and displayed the experiment results in Fig. \ref{fig:alpha_size_activation}(a). When $\alpha$ is in the interval of $[0.1,0.9]$, different $\alpha$s produce a similar identity authentication accuracy and similar average accuracy of gesture recognition. When $\alpha$ is 0.0, i.e., \textit{contrastive loss} is zero, though the identity authentication accuracy is larger than 99\%, the gesture recognition accuracies are also significant high. Some gesture recognition accuracies are even higher than 40\%. This means that the extracted feature is still highly gesture feature-sufficient. When $\alpha$ is 1.0, i.e., \textit{identity loss} has no contribution, the average accuracy of gesture recognition remains lower than 20\%. However, the identity authentication accuracy drops a lot, i.e., by approximately 10\%. We can draw two conclusions from this experiment. 1) No matter how small $\alpha$ is, as long as it is larger than 0.0, RFBP-Net can learn outstanding privacy-preserving ability. 2) During training, \textit{contrastive loss} dominates the optimization procedure. Because even when \textit{identity loss} is zero, the identity authentication accuracy is still higher than 85\%. But behavior privacy is not well protected when \textit{contrastively loss} is zero. 

\par We consider 0.8 to be the best value of $\alpha$ since the related accuracy difference is the highest one.

\subsection{Effect of feature size}
In a similar way to the RFID experiment, we tried five different feature sizes: 32, 64, 128, 256 and 512. The experiment results in Fig. \ref{fig:alpha_size_activation}(b) show that both kinds of accuracy decrease with the increase in feature size when feature size is larger than 64. When feature size is 512, both kinds of features are destroyed by RFBP-Net. Due to the fact that the overall average accuracy of gesture recognition is approximately 18\% when feature size is smaller than 512, we consider 128 to be the best choice because the related identity authentication accuracy is the highest one.

\begin{figure}[t]
    \centering
    \includegraphics[scale=0.51,trim = 50 25 65 55,clip]{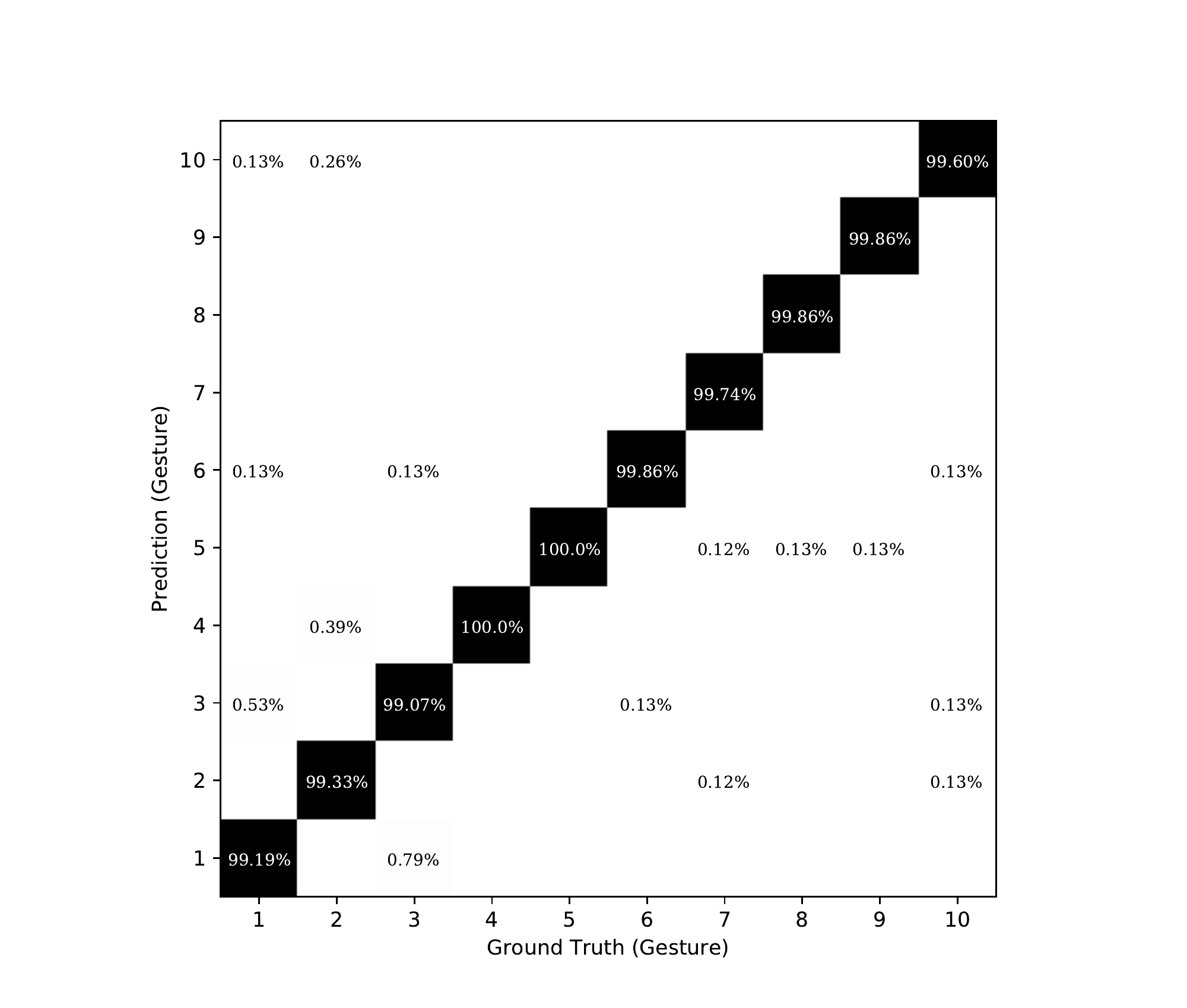}
    \caption{The confusion matrix of identity authentication.}
    \label{fig:cm_id_wifi}
\end{figure}


\subsection{Effect of activation function}
Since we think that most of the activation functions would impact the representation ability of extracted features, we did not add activation function before the output layer of RFBP-Net. In order to validate our hypothesis, we tested seven different activation functions in Fig. \ref{fig:alpha_size_activation}(c): Sigmoid \cite{DBLP:journals/ijiids/Lee14}, Tanh \cite{DBLP:conf/icdsc/AbdelouahabPB17}, ReLU \cite{DBLP:conf/fair2/PretoriusBD19}, Softplus \cite{DBLP:conf/iconip/LiuF16}, LeakyReLu \cite{DBLP:conf/icdsp/ZhangZS17}, ELU \cite{DBLP:journals/corr/ClevertUH15} and PReLU \cite{DBLP:conf/mwscas/OzawaITNO19}. The results show that by using Sigmoid and Tanh, both kinds of accuracies become significantly low. The identity authentication accuracy is even lower  than 10\%. Though behavior privacy is well protected in most of the activation functions, corresponding identity-relevant features are simultaneously destroyed. Despite the PReLU function, all the identity authentication accuracies are lower than 90\%. This phenomenon demonstrates that the majority of activation functions do diminish the representation ability of extracted features. 

\subsection{Evaluation with well-selected parameters}
According to the experiment results above, in this experiment we set training set size, $\alpha$ and feature size as 1000, 0.8 ad 128, respectively. The confusion matrix of identity authentication is shown in Fig. \ref{fig:cm_id_wifi}. The colors on the diagonal are deeper than the surrounding colors, which demonstrates that the identity-relevant feature is well retained. By contrast, the gesture recognition accuracy drops approximately 81\%, which means our framework protects behavior privacy effectively.

\begin{table}
\begin{center}
{\caption{Comparing RFBP-Net with random guess.}\label{tab:comparison_wifi}}
\begin{tabular}{ccccccc}
\cline{2-4}
\rule{0pt}{12pt}
Status&Ran. Gue.&Original Dataset&RFBP-Net\\
\hline
\\[-6pt]
\quad Identity&10\%&99\%&99\%\\
\quad Gesture&10\%&99\%&18\%\\
\hline
\end{tabular}
\end{center}
\end{table}

\begin{table*}
\begin{center}
{\caption{Evaluation with \textit{Wiar} dataset.}\label{tab:wiar}}
\begin{tabular}{cccccccc}
\cline{2-8}
\rule{0pt}{12pt}
Status&Subject 1&Subject 2&Subject 3&Subject 4&Subject 5&Average Accuracy&Identity\\
\hline
\\[-6pt]
\quad Original&91.88\%&95.63\%&96.25\%&/&/&/&/\\
\quad Unprecessed&98.32\%&96.55\%&98.00\%&95.14\%&91.38\%&95.88\%&99.71\%\\
\quad RFBP-Net&3.17\%&3.32\%&4.03\%&5.93\%&4.20\%&4.13\%&99.50\%\\
\hline
\end{tabular}
\end{center}
\end{table*}

\begin{table*}
\begin{center}
{\caption{Evaluation with \textit{Widar3.0} dataset.}\label{tab:widar}}
\begin{tabular}{cccccccccc}
\cline{2-10}
\rule{0pt}{12pt}
Status&Subject 1&Subject 2&Subject 3&Subject 4&Subject 5&Subject 6&Subject 7&Average Accuracy&Identity\\
\hline
\\[-6pt]
\quad Unprecessed&70.00\%&88.89\%&83.33\%&100.00\%&87.50\%&88.89\%&89.30\%&86.84\%&86.63\%\\
\quad RFBP-Net&15.53\%&14.75\%&7.26\%&15.67\%&13.36\%&15.75\%&17.25\%&14.22\%&71.13\%\\
\hline
\end{tabular}
\end{center}
\end{table*}

\subsection{Comparison with random guess}
In a similar way to the RFID experiment, there is no related WiFi-based previous work that can be referenced for comparison. Thus, we compare RFBP-Net with a random guess in Table \ref{tab:comparison_wifi}. It can seen that after processing by RFBP-Net, the gesture recognition accuracy approximates a random guess, yet the identity authentication accuracy is still as high as the original dataset. 

\section{Evaluation with open dataset}
\label{sec:open_dataset}
In order to further confirm the validity of RFBP-Net, we utilized our framework to process the dataset \textit{Wiar} published in \cite{DBLP:journals/access/GuoGWLLLFLSY19} and the dataset \textit{Widar3.0} published in \cite{DBLP:conf/mobisys/ZhengZ0ZLW019}. 

\subsection{Experiment with \textit{Wiar}}
\textit{Wiar} contains the WiFi signal data of ten volunteers and 16 activities. We used 2601 samples of 16 activities of five volunteers because the activity recognition accuracies of the remaining five volunteers were relatively low. The experiment results are displayed in Table \ref{tab:wiar}. In the first column, `Original' means the highest accuracy the authors of \cite{DBLP:journals/access/GuoGWLLLFLSY19} provided. `Unpressed' means the accuracy achieved by using our CNN. `RFBP-Net' means the accuracy of the data processed by RFBP-Net. The results show that RFBP-Net only causes an identity authentication accuracy reduction of 0.21\% but protects activity privacy significantly well.

\subsection{Experiment with \textit{Widar3.0}}
\textit{Widar3.0} is an open WiFi dataset published for gesture recognition study. Since \textit{Widar3.0} is a cross-domain dataset, we only used 261 samples of 6 gestures of 7 volunteers distributed in one domain. Since the number of samples is not large enough for deep learning, we use KNN, which shows best performance in NB-KNN-SVM-DT, to classify samples. The experiment results are shown in Table \ref{tab:widar}. As distinct from the negligible identity authentication accuracy reduction in \textit{Wiar}, RFBP-Net causes a 15.23\% reduction in \textit{Widar3.0}. This makes sense. since the dataset size of \textit{Wiar} is ten times that of \textit{Widar3.0}, which makes RFBP-Net stunted when we trained RFBP-Net with \textit{Widar3.0}. However, the accuracies of gesture recognition drop a lot, which demonstrates that RFBP-Net still performs well in the gesture privacy preserving of \textit{Widar3.0}.  

\begin{figure}
     \centering
     \includegraphics[scale=0.4,trim=10 2 20 20,clip]{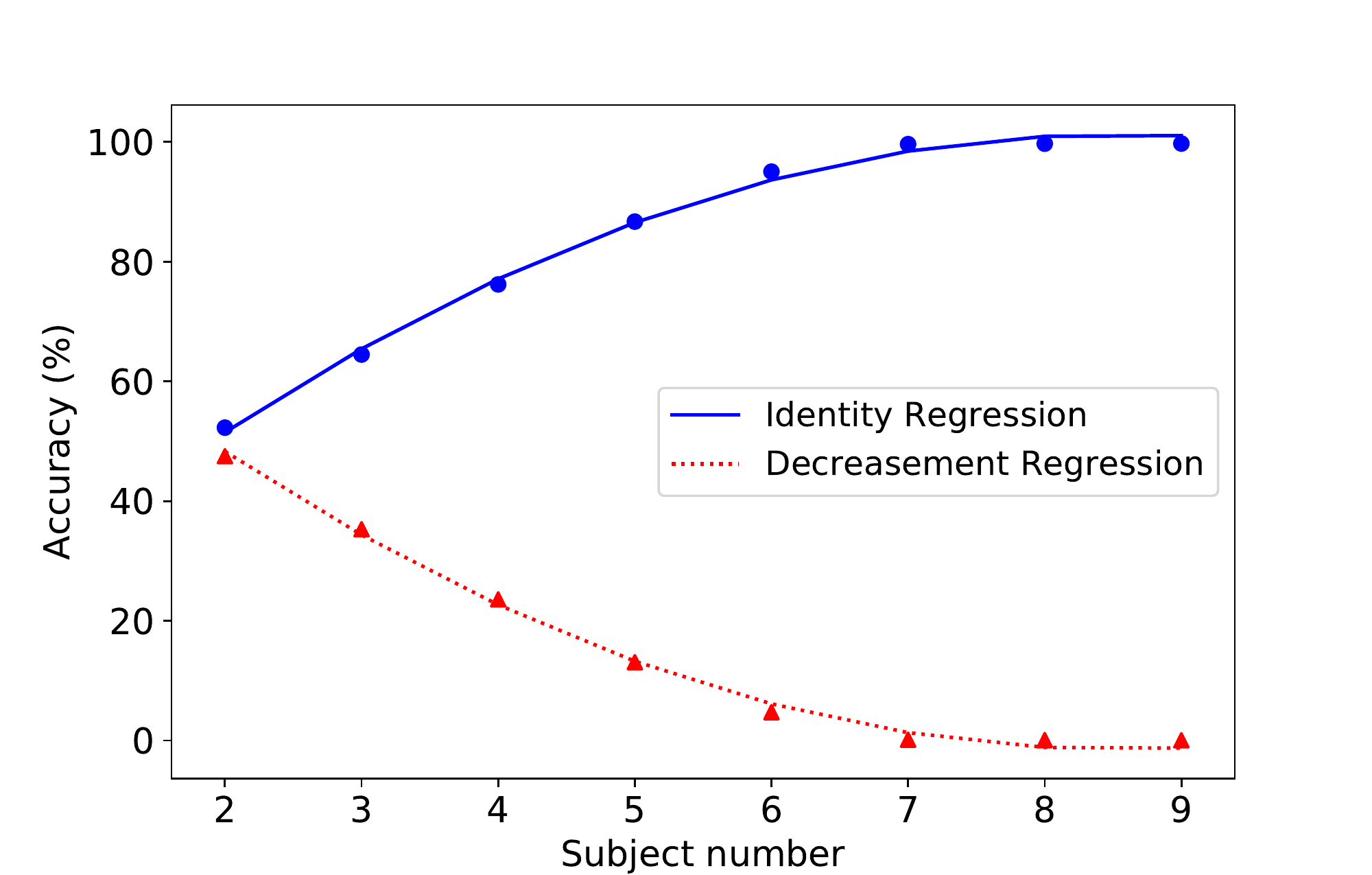}
     \caption{More subjects would yield better performance.}
     \label{fig:subject_number}
 \end{figure}

\section{Discussion and Future Work}
\label{sec:discussion}
In this section, we present two future works and a vital observation. 
\par First, in this paper, though we only utilize learning models to evaluate framework without mathematical formulas derivation, we believe that the successful experiments on five classic models (\textit{i.e.} NB, KNN, SVM, DT and NN), a deep model (\textit{i.e.}, CNN) and two open datasets (\textit{i.e.}, \textit{Wiar} and \textit{Widar3.0}) are more than enough to guarantee the validity of our framework. We will explore the formulas derivation in the future. \par Moreover, since RFBP preserving issue is different from other currently well-studied privacy-preserving issues, e.g., differential privacy preserving \cite{DBLP:journals/corr/JiLE14}, we did not use the mathematical analysis methods of other issues to analyze RFBP preserving issue. It is potential that the solutions used to protect other categories of privacy can also be leveraged to protect RFBP. We will explore this possibility in the future. 
\par Finally, it is feasible that we only invited ten volunteers (\textit{i.e.}, subjects) to participate in our signal collection, because we found that more volunteers would yield better performance. The regression results are shown in Fig. \ref{fig:subject_number}. The identity authentication accuracy of privacy-protected data increases when the number of subject increases. Meanwhile, with the increase of subject number, the reduction of identity authentication accuracy becomes smaller. Thus, we believe that RFBP-Net can perform well when the subject number is huge.   .  
 
\section{Conclusion}
\label{conclusion}
In this paper, we first defined the concept of behavior privacy in RF signal and then expressed concerns over the privacy leakage. In order to preserve RFBP in wireless human-centered applications, we propose a novel framework, whose core is RFBP-Net, for behavior-irrelevant feature extraction in user authentication system. RFBP-Net leverages a \textit{Siamese network}-based novel architecture to extract pure which can only be used for accurate identity authentication. The experiment results on a RFID system and a WiFi system showed that our framework can yield a behavior recognition accuracy of $70\%+$, trading with $5\%-$ reduction in identity authentication accuracy. 
The results of the extensive experiments on two open datasets also showed that our framework can protect behavior privacy efficiently while causing negligible reduction in identity authentication accuracy.

\bibliographystyle{ACM-Reference-Format}
\bibliography{sample-base}

\end{document}